\documentclass[aps,pre,
 onecolumn,
twocolumn,
10pt,
 notitlepage,
 showpacs,floatfix,nofootinbib,
 superscriptaddress
 ]{revtex4-2}

 \usepackage{subfigure}
 \usepackage{amssymb}
 \usepackage{amsfonts}
 \usepackage{amsmath}
 \usepackage{amsthm}
 \usepackage{epsfig}
 \usepackage{float}
 \usepackage{multirow}
 \usepackage[usenames,dvipsnames]{color}
 \usepackage[latin1]{inputenc}
 \usepackage{xr}
 \usepackage{enumitem}
 \usepackage{microtype}
 \usepackage{graphics,graphicx,xcolor,subfigure}
 \usepackage[hidelinks,unicode=true]{hyperref}
 \hypersetup{colorlinks=true,% Don't draw a box around links, instead color every piece of text, which is a link
 	linkcolor=blue,
 	urlcolor=blue,
 	citecolor=blue,
 	pdfhighlight=/N% When clicking a link and holding the mouse button, don't use any graphical effects - i.e. the alternative would be to behave like a "button" which is pressed within the PDF
 }%\usepackage{hyperref}
 
 \usepackage{comment}
 \usepackage[normalem]{ulem}

\newcommand{\av}[1]{\langle {#1} \rangle}

\newcommand{\orcid}[1]{\hspace{0.2em}\href{https://orcid.org/#1}{\includegraphics[keepaspectratio,width=0.7em]{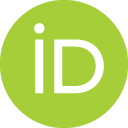}}}

\newcommand{\ra}{\rho_{\text{A}}}
\newcommand{\rb}{\rho_{\text{B}}}
\newcommand{\rab}{\rho_{\text{AB}}}
\newcommand{\rz}{\rho_{\text{0}}}
\newcommand{\rT}{\rho_\text{T}}

\newcommand{\fa}{\varphi_\text{A}}
\newcommand{\fb}{\varphi_\text{A}}

\newcommand{\bvphi}{\bar{\varphi}}

\newcommand{\rak}{\rho^{\text{A}}_k}
\newcommand{\rbk}{\rho^{\text{B}}_k}
\newcommand{\rabk}{\rho^{\text{AB}}_k}
\newcommand{\rzk}{\rho^{\text{0}}_k}
\newcommand{\rk}{\rho_k}

\newcommand{\fak}{\varphi^\text{A}_k}
\newcommand{\fbk}{\varphi^\text{B}_k}
\newcommand{\rTk}{\rho^\text{T}_k}

\newcommand{\brabk}{\bar{\rho}^{\text{AB}}_k}

\newcommand{\brk}{\bar{\rho}_k}

\newcommand{\fakl}{\varphi^\text{A}_{k'}}
\newcommand{\fbkl}{\varphi^\text{B}_{k'}}
\newcommand{\rTkl}{\rho^\text{T}_{k'}}

\newcommand{\kmax}{k_\text{max}}
\newcommand{\kmin}{k_\text{min}}
\newcommand{\mus}{\mu_\text{s}}

\graphicspath{{./}}

\begin{document}

\title{Heterogeneous mean-field theory for two-species symbiotic processes on networks}

\author{Guilherme S. Costa\orcid{0000-0002-5019-0098}}
\affiliation{Departamento de F\'{\i}sica, Universidade Federal de Vi\c{c}osa, 36570-900 Vi\c{c}osa, Minas Gerais, Brazil}

\author{Marcelo M. de Oliveira \orcid{0000-0002-8940-1972}}
\affiliation{Departamento de Estat\'{\i}stica, F\'{\i}sica e Matem\'atica, Universidade Federal de S\~ao Jo\~ao del-Rei, 36420-000, Ouro Branco , Minas Gerais, Brazil}

\author{Silvio C. Ferreira\orcid{0000-0001-7159-2769}}
%\email{silviojr@ufv.br}
%\thanks{Corresponding author.}
\affiliation{Departamento de F\'{\i}sica, Universidade Federal de Vi\c{c}osa, 36570-900 Vi\c{c}osa, Minas Gerais, Brazil}
\affiliation{National Institute of Science and Technology for Complex Systems, 22290-180, Rio de Janeiro, Brazil}

\begin{abstract}
A simple model to study cooperation is the two-species symbiotic contact process (2SCP), in which two different species spread on a graph and interact by a reduced death rate if both occupy the same vertex, representing a symbiotic interaction. The 2SCP is known to exhibit a complex behavior with a rich phase diagram, including  continuous and discontinuous transitions between the active phase and extinction. In this work, we advance the understanding of the phase transition of the 2SCP on uncorrelated networks by developing a heterogeneous mean-field (HMF) theory,  in which the heterogeneity of contacts is explicitly reckoned. The HMF theory for networks with power-law degree distribution shows that the region of bistability (active and inactive phases) in the phase diagram shrinks  as the heterogeneity level is increased by reducing the degree exponent. Finite-size analysis reveals a complex behavior where a pseudo discontinuous transition at a finite-size can be converted into a continuous one in the thermodynamic limit, depending on degree exponent and symbiotic coupling.  The theoretical results are supported by extensive numerical simulations.
\end{abstract}
 %\pacs{}
	
\maketitle

\section{Introduction}

Cooperative or symbiotic processes, in which two or more dynamics evolve on the same substrate, interacting with each other synergistically, present a rich behavior in terms of the phase transitions and criticality~\cite{Wang2019}. This kind of approach is extensively used in ecological models of competition and cooperation~\cite{Iwata2011,Ulrich2015}. However, this idea can be expanded to the context of epidemics if one considers two interacting pathogens propagating across the same hosts.  If a host can be infected simultaneously by both pathogens, the co-infection can result in coexisting infections when competitive interactions are considered~\cite{Newman2005,Newman2013}. An important class of interacting dynamical processes are those involving information and epidemic spreading influencing each other~\cite{Bianconi2021,Granell2013,Wang2016,Bedson2021}.  On the other hand, cooperative or synergistic interactions result in richer phase diagrams, which may include discontinuous phase transitions and they have been a topic of intense research~\cite{Newman2013, Chen2013, Cai2015, Grassberger2016, Janssen2016, Cui2017, Liu, Baek2019, deOliveira2012, deOliveira2019}. 

Phase transitions in spreading phenomena involve the absorbing states~\cite{Pastor-Satorras2015,Castellano2009},  which are the frozen configurations without fluctuations of the order parameter~\cite{Marro2005}.  Two fundamental models for spreading of single species are the susceptible-infected-susceptible (SIS) epidemic model~\cite{Pastor-Satorras2015} and the contact process (CP)~\cite{Harris1974,Castellano2006}. These models on generic graphs are defined as follows. Nodes can be active (infectious) or inactive (susceptible).  Active nodes become spontaneously inactive with rate $\mu$, in both models, or can independently activate each inactive nearest neighbor with rates $\lambda$ and $\lambda/k$ in SIS and CP models,{respectively}, in which $k$ is the node degree (number of neighbors). Despite the similar rules, {these} processes behave very differently on heterogeneous networks. Considering the important case of power-law degree distributions, in which the probability that a randomly chosen node has degree $k$ {scales} as $P(k)\sim k^{-\gamma}$~\cite{Albert2002} where $\gamma$ is the degree exponent, CP presents a finite activation threshold~\cite{Ferreira2011,Mata2014} and a phase transition  while the SIS is governed by complex activation mechanisms~\cite{Ferreira2016a} which leads to vanishing epidemic threshold, and consequently absence of a genuine phase transition in the thermodynamic limit~\cite{Chatterjee2009,Boguna2013}.

While coexisting spreading processes on networks are widely investigated using models based on SIS-like dynamics~\cite{Granell2013,Wang2016,Sanz2014}, the phase transition can be more naturally tackled in CP-like spreading processes where {the} phase transitions happen in the thermodynamic limit. A simple model to study cooperation is the two-species symbiotic CP (2SCP)~\cite{deOliveira2012}, in which two different species spread on a substrate following the standard CP rules except if both occupy the same node, when they  interact symbiotically by a reduced death rate $\mus$. In addition to its interest as an elementary model of symbiosis, the 2SCP is useful in the study of out-of-equilibrium phase transitions. Several works discussing and characterizing the phase transition of this model were out recently~\cite{deOliveira2012, deOliveira2014, Sampaio2018, deOliveira2019}. On regular lattices, the 2SCP presents a continuous phase transition in one and two spatial dimensions~\cite{deOliveira2012}. However, it was found that the transition becomes discontinuous in the regime of strong symbiosis when diffusion is introduced~\cite{deOliveira2014}. The 2SCP was also investigated in complete graphs and random regular (RR) networks~\cite{Sampaio2018}, and it was conjectured that the nature of its transition changes, from continuous to discontinuous, at the upper critical dimension. The phase diagram determining the regions of the 2SCP space parameter $\mus$ versus $\lambda$ was obtained in the simplest one-site mean-field level~\cite{deOliveira2014}. 

The 2SCP was {also} investigated numerically in complex networks (Barabási-Albert, Erdös-Renyi and RR networks) in Ref.~\cite{deOliveira2019} and the results compared with a homogeneous pairwise mean-field theory where the fixed degree is replaced with the average degree of the network. This strategy was  previously used for ordinary CP on networks~\cite{Ferreira2011,Juhasz2012}. An approach to reckon the heterogeneity explicitly is the heterogeneous mean-field (HMF) theory, conceived to investigate dynamical processes on complex networks~\cite{Pastor-Satorras2015}. It assumes that the vertex degree is the  quantity relevant to determine its state, neglecting dynamical correlations as well as the actual structure of the network. While failing to reproduce accurately the activation of the SIS dynamics on {power-law networks} with $\gamma>5/2$ due to strong localization effects~\cite{Boguna2013,Ferreira2016a}, it reproduces very accurately the CP critical behavior on these same networks~\cite{Ferreira2011,Mata2014}. In order to contribute to the better understanding of symbiotic dynamics on complex networks, we develop a HMF theory~{\cite{Pastor-Satorras2001,Pastor-Satorras2015}} for 2SCP and analyze the case of power-law networks considering {a} range of  degree exponent $2<\gamma<4$.  We validate the results by performing quasistationary (QS) simulations on synthetic complex networks. We report that the degree distribution plays a central role in the shape of the phase diagram defining active, inactive, and bistable phases, in which the last one shrinks for lower degree exponents (higher heterogeneity). Finite-size scaling reveals complex behaviors where a pseudo discontinuous transition at finite sizes becomes continuous in the thermodynamic limit.

{This} paper is organized as follows. In Section~\ref{sec:modelHMF}, we present the model, review  basic properties for homogeneous networks and develop the HMF theory for S2CP. In Sec.~\ref{sec:numerics} we perform the numerical analysis of the HMF equations and finite-size scaling and stochastic simulations are compared with the HMF theory in Sec.~\ref{sec:simu}. Finally, Sec.~\ref{sec:conclu} is devoted to summarizing our conclusions.

\section{Mean-field theories for the 2SCP}
\label{sec:modelHMF}

\begin{figure}[ht]
	\centering
	\includegraphics[width=0.8\linewidth]{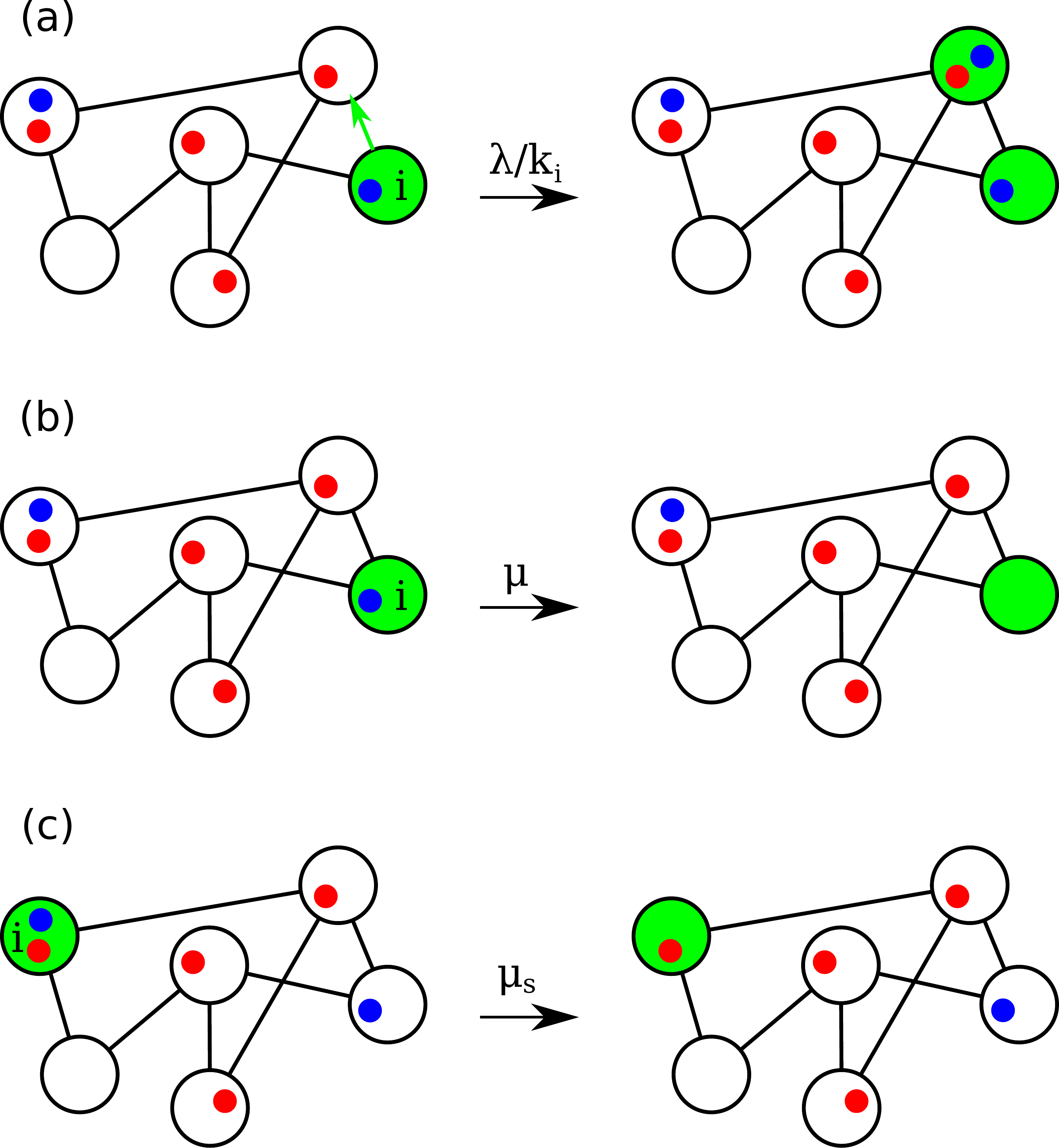}
	\caption{Transitions of the 2SCP model on networks where species are represented by red and blue dots while the nodes involved in the transition are depicted in green. (a) Individuals of species A or B  in the node $i$ replicate a copy of {themselves} in a neighbor node if allowed (doubly occupation with the same species is forbidden). Spontaneous deaths on nodes  that contain (b) one and (c) both species (symbiosis) happen with rates $\mu$ and $\mus$, respectively.}
	\label{fig:2scpmod}
\end{figure}

The 2SCP is defined considering two species (A and B) evolving on the same network. Each node can {support} at most one individual of each species. The activation process is exactly the same of the CP where both species create clones of themselves with rate $\lambda/k$ at all neighbors that do not carry one individual of its own species. {In addition, a} singly occupied node by either A or B becomes vacant with rate $\mu$. If a node contains two species, a reduced symbiotic death rate $\mus < \mu$ is adopted, such that the chance of death for both A and B individuals is reduced. All transitions for the model are illustrated in Fig.~\ref{fig:2scpmod}. Hereafter, we adopt $\mu = 1$ without loss of generality.

\subsection{Homogeneous mean field}
\label{sub:MF}
We start reviewing the basic homogeneous mean-field theory for 2SCP~\cite{deOliveira2012}, where all nodes are statistically equivalent. Therefore, the dynamical equations are constructed in terms of probabilities that the node is in a given state: vacant, occupied  by one species A, by one species B, or by both species. Taking into account all possible transitions indicated in Fig.~\ref{fig:2scpmod}, the following set of equations is obtained
\begin{align}
\dfrac{d \rz}{dt} &= \ra + \rb - \lambda \rz \rT, \\
\dfrac{d \ra}{dt} &= -\ra + \rab \mus + \lambda \rz\fa - \lambda \ra \fb, \\
\dfrac{d \rb}{dt} &= -\rb + \rab \mus + \lambda \rz\fb - \lambda \rb \fa, \\
\dfrac{d \rab}{dt}&= - 2\mus \rab + 2\lambda \ra \rb + \lambda \rab (\ra + \rb), 
\end{align}
in which $\rz$, $\ra$, $\rb$ and $\rab$ are probabilities that a given node is vacant, occupied by an individual of species $A$, of species $B$, or both, respectively. {The auxiliary variables $\rT = \ra + \rb + 2\rab$ is the total prevalence and $\varphi_\text{X}=\rab+\rho_\text{X}$ is the probability that a node has at least one individual of species X.} It is important to note that in the absence of either $A$ or $B$, the {mean-field} equations reduce to {the ones of the} standard CP~\cite{Marro2005}. By considering symmetrical solutions $\ra = \rb = \rho$ and the closure relation $\ra + \rb + \rab + \rz = 1$, the system is reduced to two independent variables.
 
The stationary solutions  are the trivial $\bar{\rho}=\bar{\rho}_{\text{AB}}=0$ and the nontrivial one given by 
\begin{equation}
\bar{\rho} = \dfrac{\mus\left[ 2(1-\mus) - \lambda + \sqrt{\lambda^2 - 4\mus (1-\mus)}\right]}{2 \lambda(1 - \mus)} 
\label{eq:homo2scp}
\end{equation} 
and
\begin{equation}
\bar{\rho}_{\text{AB}} = \dfrac{ \lambda \bar{\rho}^2}{\mus - \lambda \bar{\rho}}.
\end{equation}
 
{Analyzing these solutions, the} following conclusions can be obtained: (i) The nontrivial solution exists only if $\lambda \geq \sqrt{4\mus(1-\mus)}$; (ii) For $\mus > 1/2$, $\lambda_\text{c}=1$ is a continuous transition point where the scaling $\bar{\rho}\simeq \tfrac{\mu}{2\mu-1}(\lambda-\lambda_\text{c})$ holds;  (iii) For $\mus<1/2$, the transition point $\lambda_\text{c} = \sqrt{4\mus(1-\mus)}$ implies in $\bar{\rho} > 0$ for the nontrivial solution, indicating a discontinuous transition.

The discontinuity in solutions for $\mus < 1/2$ implies in a bistability region since $\rho=0$ is also locally stable for $\sqrt{4\mus(1-\mus)}<\lambda<1$. The convergence to the stationary state depends on the initial condition: for {$\ra(0)=\rb(0) \lessapprox 1$}, the dynamics converges to Eq.~\eqref{eq:homo2scp}  for $\lambda > \lambda^- =  \sqrt{4\mus(1-\mus)}$ while for {$\ra(0)=\rb(0) \gtrapprox 0$}, the convergence happens for $\lambda> \lambda^+ = 1$  while the absorbing state remains stable otherwise. The curves $\lambda^-(\mus)$ and $\lambda^+(\mus)$ are called lower and upper {spinoidals}, respectively. Visual representations of {the} discontinuity {are} shown in  Fig.~\ref{fig:phasediagram}.{Although it is} beyond the scope of the paper, it is important to mention that homogeneous theories were constructed using pairwise interactions~\cite{deOliveira2019}.

\subsection{Heterogeneous mean-field}
\label{sub:HMF}

\begin{figure*}[hbt]
\centering
\includegraphics[width=0.35\linewidth]{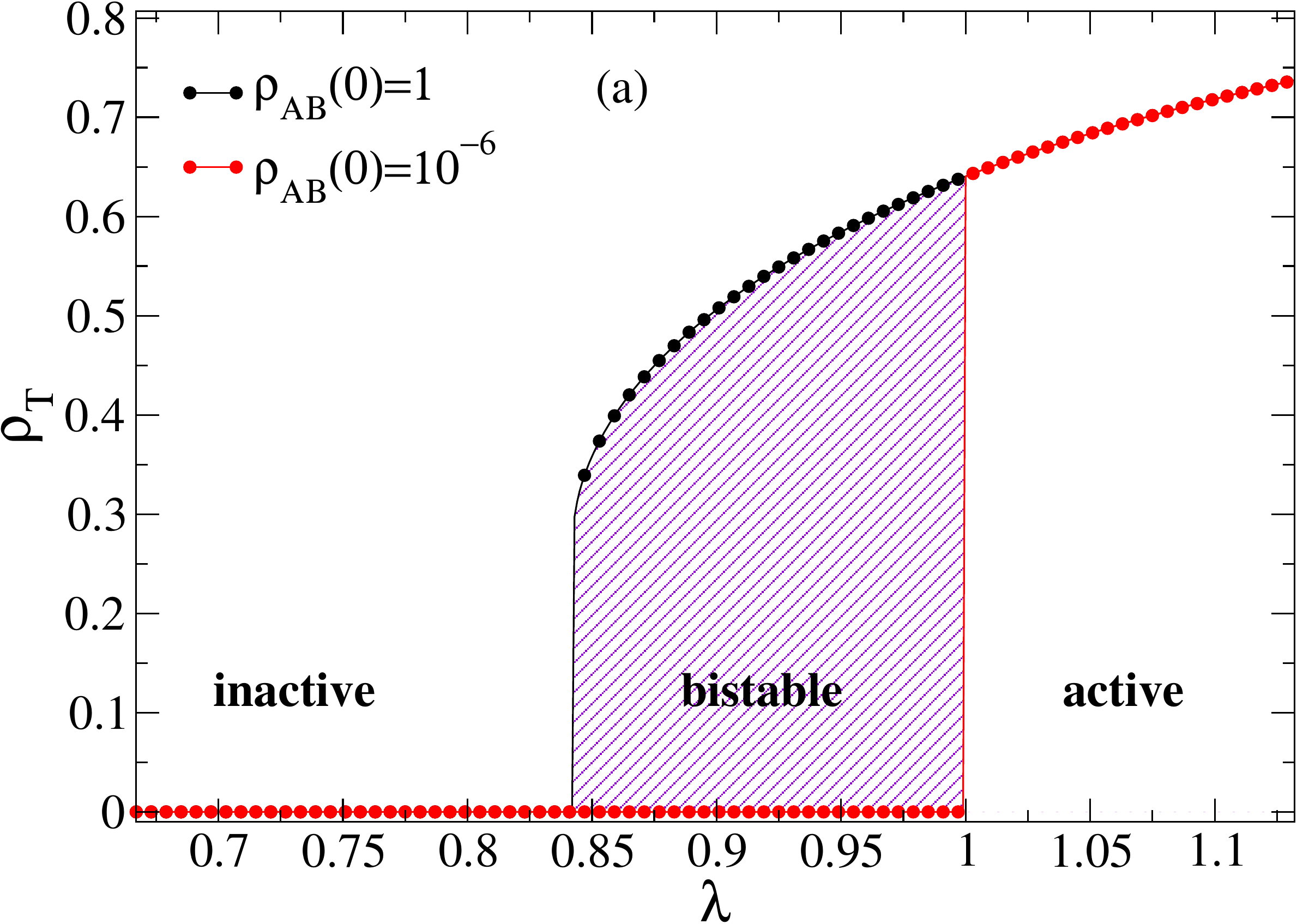}
\includegraphics[width=0.35\linewidth]{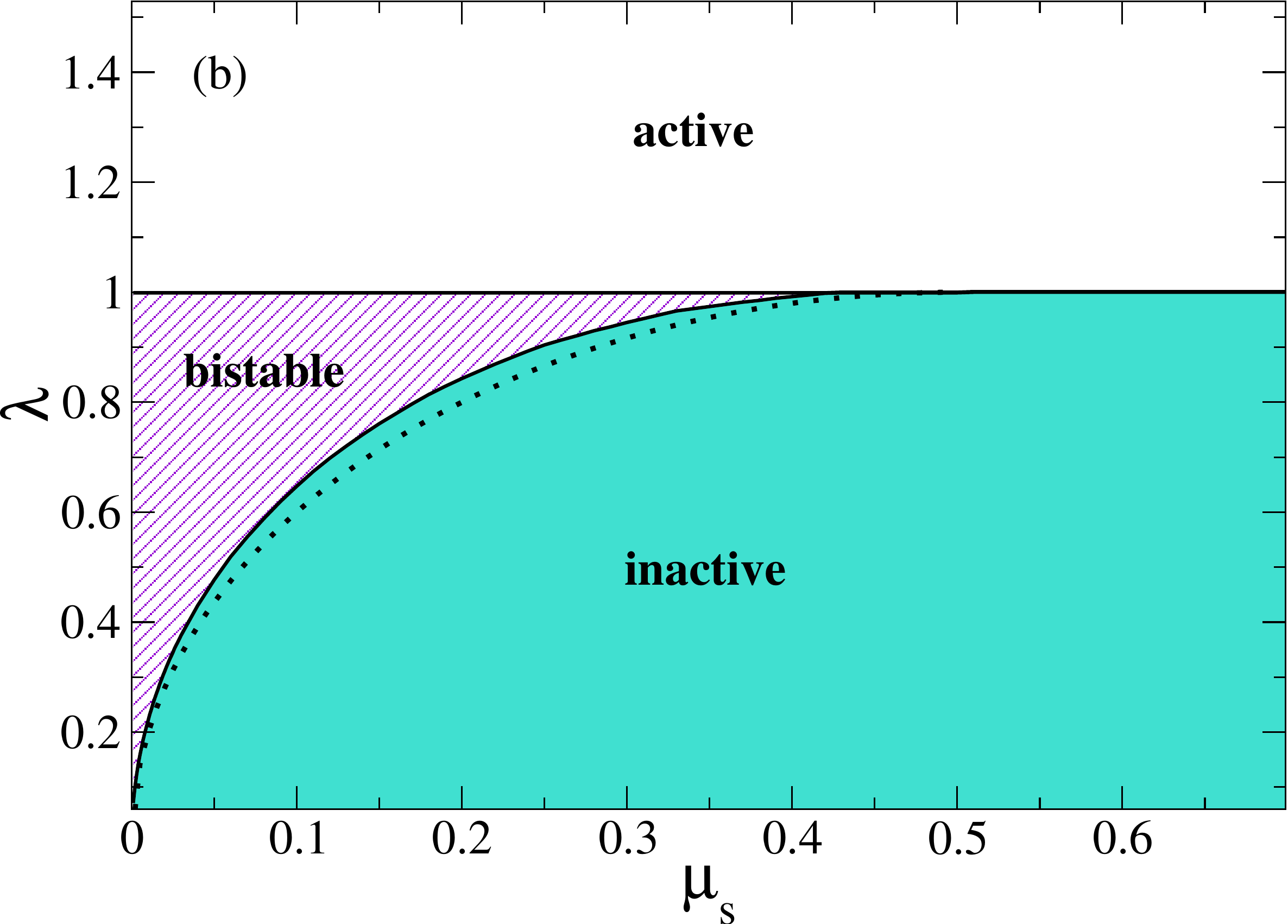}\\
\includegraphics[width=0.35\linewidth]{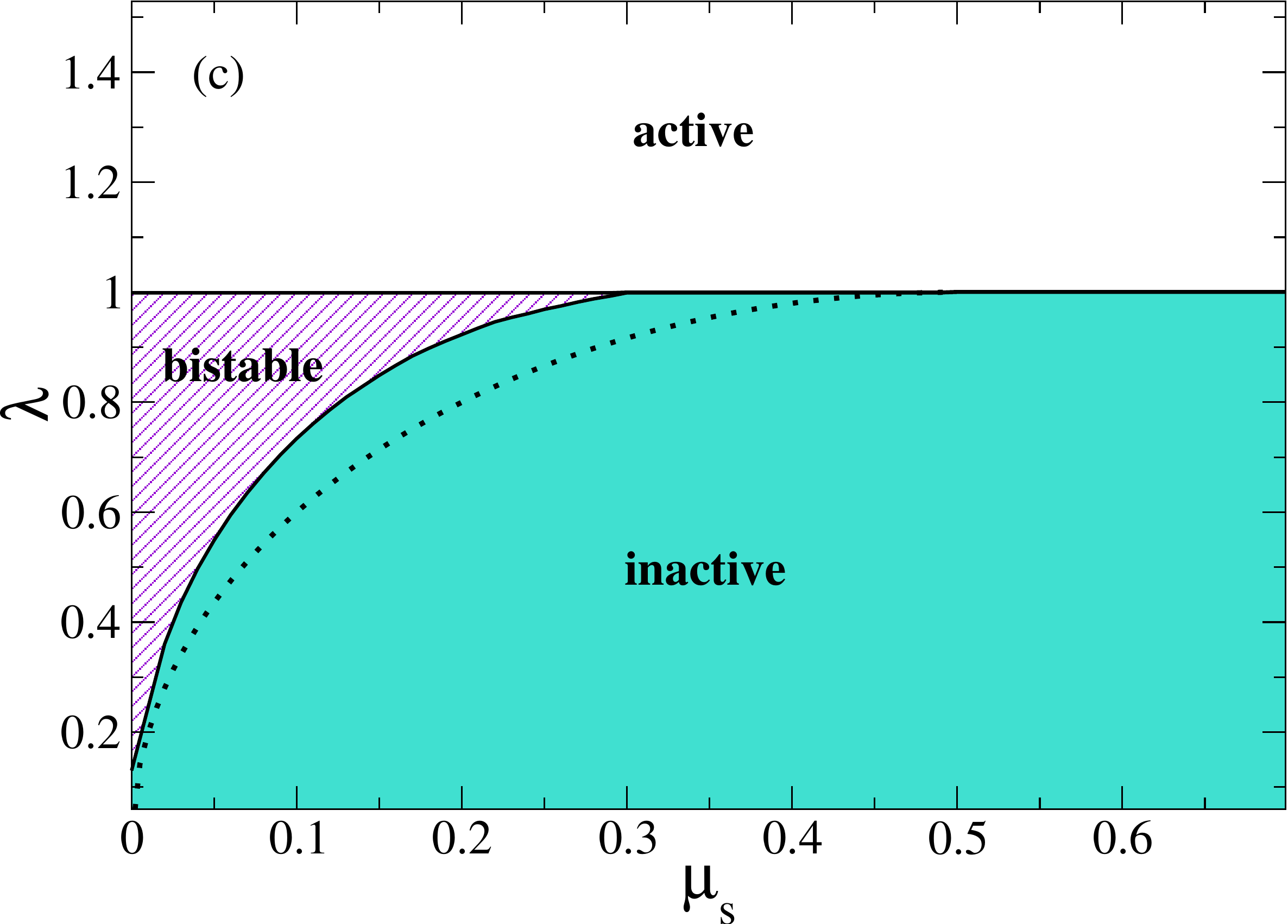}
\includegraphics[width=0.35\linewidth]{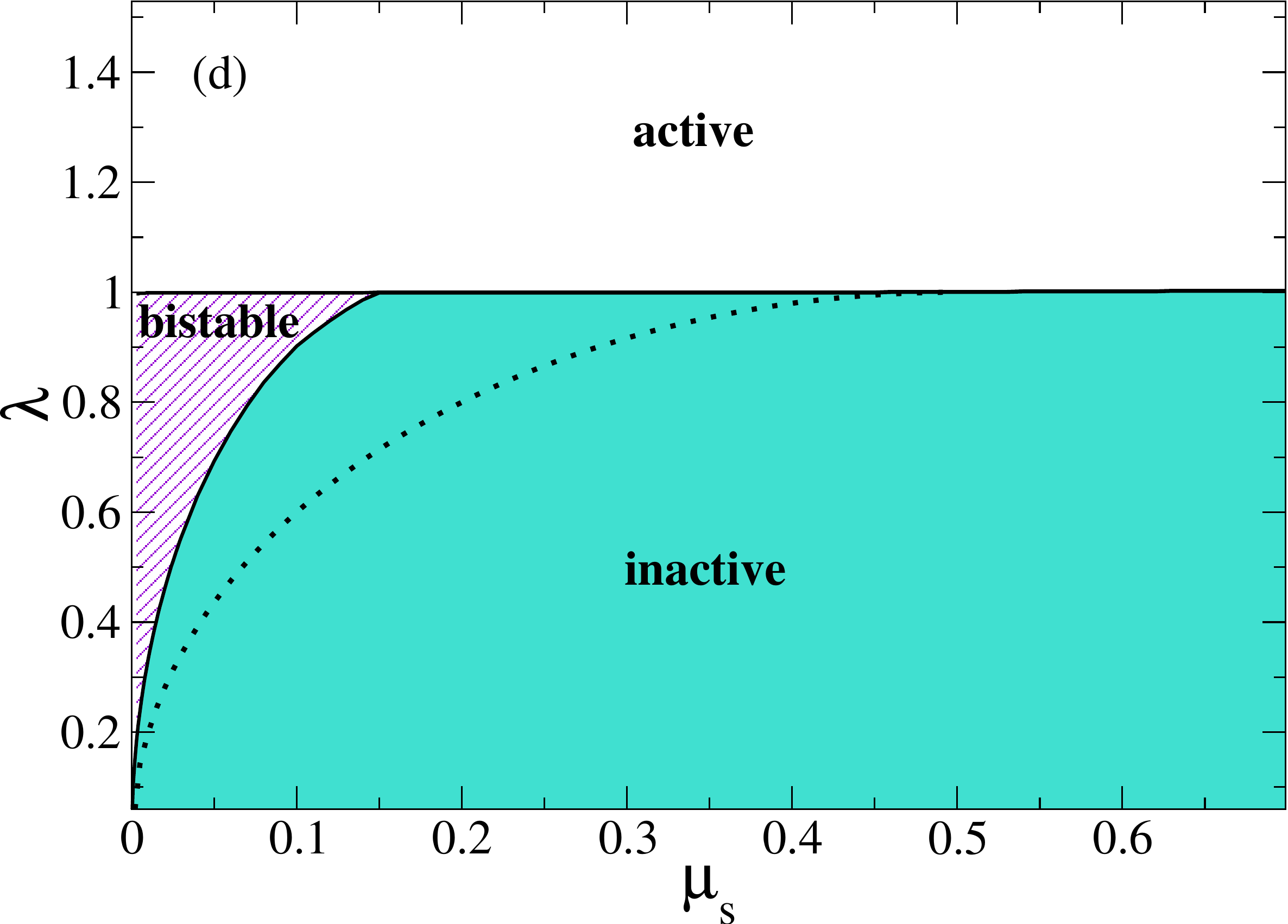}
\caption{HMF theory for the 2SCP model for a power-law degree distributions with lower and upper cutoffs given by $\kmin=3$ and $\kmax=17320$, respectively, and degree exponent $\gamma$. (a) Example of hysteresis in the total prevalence curves for $\gamma = 3.5$ and $\mus = 0.2$. The hashed area indicates the bistability region. Phase diagram for 2SCP considering (b) $\gamma = 3.5$, (c) $\gamma=2.7$, and (d) $\gamma=2.3$. The dotted lines represent the lower spinodal for the homogeneous theory $\lambda^{-}=\sqrt{4\mus(1-\mus)}$.}
	\label{fig:phasediagram}
\end{figure*}

The basic assumptions of the HMF theory~{\cite{Pastor-Satorras2001,Pastor-Satorras2015}} is that all nodes of same degree are equivalent and interact with other nodes considering the conditional probability $P(k'|k)$ that a node of degree $k$ is connected to another node of degree $k'$. We introduce the notation  $\rho_k^{\text{X}}$  for the  prevalence of individuals of  species X  (A, B, vacant, or AB) lying on nodes of degree $k$, i.e., the probability that a randomly chosen  node of degree $k$ is in the state X. {We also extend the definitions of the auxiliary variables $\fak$, $\fbk$, and $\rTk$ to their degree dependence.} Thus, dynamical equations for the HMF theory become
\begin{equation}
\dfrac{d \rho_k^0}{dt} = \rho_k^A +\rho_k^B - \lambda \rho_k^0 k \sum_{k'} \dfrac{\rTkl P(k'|k)}{k'}, 
\end{equation}
\begin{eqnarray}
\dfrac{d\rho_k^A}{dt}  & = & -\rho_k^A + \mus \rho_k^{AB}  + \lambda \rho_k^0 k \sum_{k'}  \dfrac{\fakl P(k'|k)}{k'}  \nonumber \\ 
& & -\lambda \rho_k^A k \sum_{k'}  \dfrac{\fbkl P(k'|k)}{k'}, 
\end{eqnarray}
\begin{eqnarray}
\dfrac{d\rho_k^B}{dt} & = & -\rho_k^B + \mus \rho_k^{AB}  + \lambda \rho_k^0 k \sum_{k'}  \dfrac{\fbkl P(k'|k)}{k'} \nonumber\\
& & -\lambda \rho_k^B k \sum_{k'}  \dfrac{\fakl P(k'|k)}{k'},
\end{eqnarray}
and
\begin{eqnarray}
\dfrac{d\rho_k^{AB}}{dt} &=& -2\mus\rho_k^{AB} + \lambda \rho_k^{A} k \sum_{k'} \dfrac{\fbkl P(k'|k)}{k'}  \nonumber \\
& &+\lambda \rho_k^{B} k \sum_{k'} \dfrac{\fakl P(k'|k)}{k'}.
\end{eqnarray}

As in the homogeneous theory, we look for symmetric ($\rak = \rbk = \rk$) and stationary solutions ($\dot{\rho}^\text{X} = 0$) and can use the closure relation  $\rzk + \rak + \rbk + \rabk = 1$. In order to advance in the solution, we consider uncorrelated networks for which the neighbor's degree is independent of the degree of the considered node resulting in $P(k'|k) = {k'P(k')}/{\langle k \rangle}$~\cite{barabasi2016network}. Lets define $\beta_k = {\lambda k}/{\langle k \rangle}$ and {$\varphi = \sum_{k}\fak P(k)= \sum_{k}\fbk P(k)$}, in which the latter does not depend on $k$,  to obtain  the following set of equations for $k=\kmin,\ldots,\kmax$ 
\begin{equation}
\frac{d\rho_k}{dt} {=} -{\rho}_k +  \mus\rabk + (1-3{\rho}_k-\rabk) \beta_k {\varphi} 
\label{eq:rhok}
\end{equation}
and
\begin{equation}
\frac{d\rabk}{dt} = - \mus \rabk + \beta_k {\rho}_k {\varphi}.
\end{equation}

Solving these equations in the steady state for $\rabk$ and $\rk$, one obtains
\begin{equation}
\brabk = {\beta_k \rho_k \bvphi}/{\mus} 
\label{eq:rhoab_k}
\end{equation}
and
\begin{equation}
\brk = \dfrac{\beta_k \bvphi }{1+2\beta_k \bvphi + {\beta_k^2 \bvphi^2}/{\mus}}.
\label{eq:rho_k}
\end{equation}
Substituting Eqs.~\eqref{eq:rhoab_k} and \eqref{eq:rhok} in $\bvphi$, we obtain a self-consistent transcendent equation
\begin{equation}
\bvphi = \Theta(\bvphi) = \sum_{k}\dfrac{\bvphi \beta_k  (1+\beta_k\bvphi/\mus)   P(k)}{1+2\beta_k \bvphi + \beta_k^2 \bvphi^2/{\mus}},
\label{eq:varphi}
\end{equation}
which can be expanded in terms of Gauss hypergeometric functions~\cite{zwillinger2007table} using a continuous degree approximation (Appendix~\ref{app:HMF}) or solved numerically using bisection method given a degree distribution $P(k)$ and, thus, providing $\brk$ and $\brabk$. 

The loss of stability of the absorbing state $\rk=\rabk=0$ at $\lambda^+$ can be obtained using Eq.~\eqref{eq:rhoab_k}  near to the transition point, where both  $\rk\ll 1$ and $\rabk\ll 1${, implying that $\bvphi\ll 1$}. According to Eq.~\eqref{eq:rhoab_k}, {which is proportional to the product $\rho_k \bvphi$,} we can assume $\rabk\ll \rk$ in Eq.~\eqref{eq:rho_k} to obtain the following linearized and closed system for $\rk$
\begin{equation}
\frac{d\rho_k}{dt} = -\rho_k+\frac{\lambda k}{\av{k}}\sum_{k'}\rho_{k'}P(k')+\ldots\simeq \sum_{k'} L_{kk'}\rho_{k'},
\end{equation}
where  
\begin{equation}
L_{kk'}=-\delta_{kk'}+ \frac{\lambda kP(k')}{\av{k}}
\end{equation}
is the Jacobian of the linearized system. This is exactly the same Jacobian of the single species CP on networks~\cite{Boguna2009}. The loss of stability of the absorbing state is obtained when the largest eigenvalue of $L_{kk'}$ is zero. {One can easily check that $v_k=k$ is an eigenvector of $C_{kk'} = kP(k')/\av{k}$ with eigenvalue $\Lambda=1$. Since $v_k>0$ and $C_{kk'}$ is positive definite, Perron-Frobenius theorem guaranties that $\Lambda$ is the largest and non-degenerate eigenvalue of $C_{kk'}$,} providing the upper spinodal $\lambda^{+}=1$.

We define the order parameter as the fraction of nodes occupied by at least one particle, which is given by
\begin{equation}
\rho_\text{T} = 2\rho + \rho_{\text{AB}},
\end{equation}
in which
\begin{equation}
\rho = \sum_{k=\kmin}^{\kmax} \rho_k P(k)
\end{equation}
{is} the probability that a randomly chosen node is occupied by either A or B species and 
\begin{equation}
\rho_{\text{AB}} = \sum_{k=\kmin}^{\kmax} \rabk P(k)
\end{equation}
is the probability  of double occupation. In figure \ref{fig:phasediagram}(a) we present typical hysteresis diagrams for $\rho_\text{T}$ as a function of $\lambda$ using a power-law degree distribution with $\gamma = 3.5$ and $\mus = 0.2$ for two initial conditions: $\rho_{\text{AB}}(0) = 1$ representing a fully occupied substrate and $\rho_{\text{AB}}(0) = 10^{-6}$ which is near to the absorbing state. To compare {this result} with the  {case of}  uncorrelated networks in Sec.~\ref{sec:simu}, we have chosen an upper cutoff for the degree distribution $\kmax=\sqrt{N}$, where $N$ is the number of nodes of the network. The hysteresis effect in these curves manifests as bistability regions (hashed area). The phase diagrams in the $\lambda \times \mus$ parameter space computed numerically solving Eq.~\eqref{eq:varphi} are presented for degree exponents $\gamma=3.5$, $2.7$ and $2.3$ in Fig.~\ref{fig:phasediagram}. The upper spinodal   $\lambda^+=1$ is confirmed.  One can see that the bistability region is close to the homogeneous case for $\gamma=3.5$, being gradually reduced as the network heterogeneity is increased with smaller values of $\gamma$, shrinking in the limit $\gamma\rightarrow 2$. 

One can rationalize the role of heterogeneity as follows. The chance of a doubly occupied node to produce occupation with both species in one of its neighbors decreases with its degree due to the random choice of the target, scaling approximately as $1/k^2$ (choose the same neighbor twice for both A and B offspring before death). So, even though hubs are, on average, more active than the regular nodes, more hubs dismantle the symbiotic mechanisms by diluting species in different neighbors.

%?For \gamma = 3.5, it is very close to the homogeneous network case, \mu_s = 1/2, shown in Sec. II A.? Add a paragraph break before ?We then ?? below and change ?We then ?? to ?We ??

The {diagrams} for finite-size system  {indicate} that the transitions become continuous for $\mus>\mus^*$, which depends on the degree exponent $\gamma$. For $\gamma=3.5$, it is very close to homogeneous network case{,} $\mus^*=1/2${,} shown in Sec.~\ref{sub:MF}. 

\begin{figure}[tbh]
	\centering
	\includegraphics[width=0.9\linewidth]{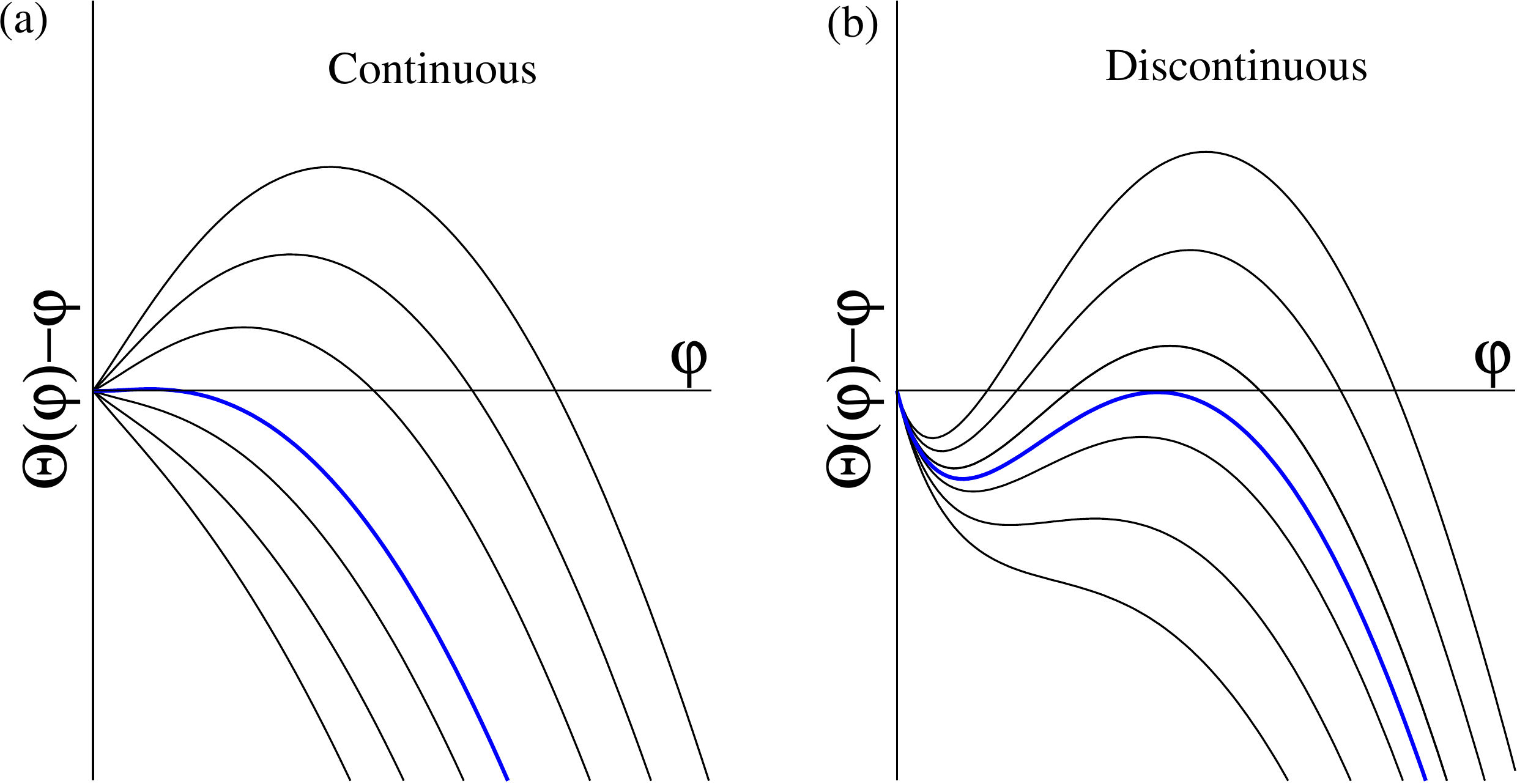}
	\caption{{Example} of a (a) continuous and (b) discontinuous transition through transcritical and pitchfork bifurcations, respectively. {The curves correspond to the solution of Eq.~\eqref{eq:varphi} with (a) $\mus=0.3$ and (b) $\mus=0.05$ for a degree exponent $\gamma=2.7$. The contagion rate $\lambda$ is increased from bottom to top.} The blue curves are the critical ones.}
	\label{fig:scheme}
\end{figure}

We perform a continuous approximation~\cite{Mata2014} for Eq.~\eqref{eq:varphi}, where the sum is replaced by an integral over $k$, to obtain 
\begin{equation}
\bvphi=\Theta(\bvphi) = \int_{\kmin}^{\infty} \dfrac{\bvphi \beta_k  (1+\beta_k\bvphi/\mus)   P(k)}{1+2\beta_k \bvphi + \beta_k^2 \bvphi^2/{\mus}}dk
\label{eq:varphi_cont}
\end{equation}
in the limit $\kmax\rightarrow \infty$. One can verify by direct differentiation that $\Theta'(\varphi)>0$ and $\Theta''(\varphi)<0$ for $\mus\ge 1/2$, implying that $\Theta(\varphi)$ is monotonically increasing function and that only a continuous transition is possible, as illustrated in Fig.~\ref{fig:scheme}(a). Therefore, a discontinuous transition {shown in Fig.~\ref{fig:scheme} (b)}, if there is one, must occur for $\mus\le 1/2$. To determine when the discontinuous transition turns to continuous, i.e., when the gap in lower spinodal vanishes, we can use a series expansion for small $\bvphi$. 

\begin{figure}[tbh]
	\centering
	\includegraphics[width=0.8\linewidth]{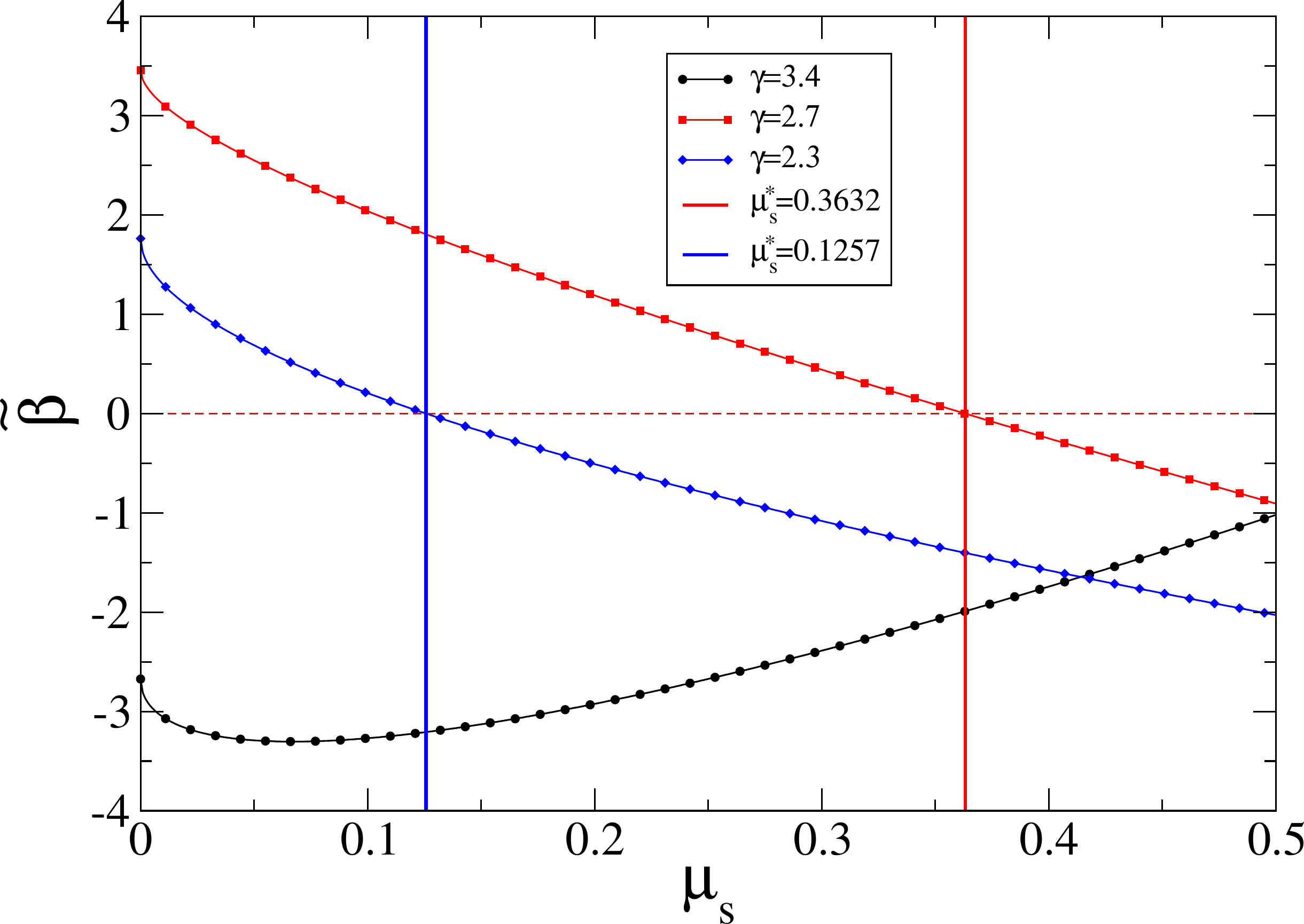}
	\caption{Coefficient $\tilde{\beta}$, where $a_{\gamma-1}\propto\tilde{\beta}$, as function of symbiotic parameter $\mus$ on power-law networks with different values of the degree exponent $\gamma$ indicated in the legend. Vertical lines are roots of $\tilde{\beta}(\mus)=0$ for $\gamma=2.7$ and $2.3$.}
	\label{fig:betatil}
\end{figure}

A lengthy algebraic handling, summarized in Appendix~\ref{app:HMF}, leads to
\begin{equation}
\Theta(\bvphi) = \lambda\bvphi+a_{\gamma-1}\bvphi^{\gamma-1}+a_2\bvphi^2+a_3\phi^3,
\end{equation}
where the coefficients $a_x$ are functions of $\mus$, $\gamma$, and $\lambda$ given in Eq.~\eqref{eq:Theta_series}. Considering the range of interest $\mus\le 1/2$, we can easily see that $a_2\le 0$ for $2<\gamma<3$ and $a_2\ge 0$ for $\gamma>3$. Similarly, $a_3>0$ for $\gamma<4$ and $a_3<0$ for $\gamma>4$. The coefficient $a_{\gamma-1}$ 
has a very complicated dependence on $\gamma$ and $\mu_s $ shown in Eq.~\eqref{eq:a_gamma_m_1}. However, the sign of $a_{\gamma-1}$ can be investigated numerically. We found that $a_{\gamma-1}<0$ for $\gamma>3$ while changes sign in interval $0\le\mus\le1/2$ for $2<\gamma<3$; see Fig.~\ref{fig:betatil}.

The cubic term is negligible for $\gamma<4$ while the term $\bvphi^{\gamma-1}$ is negligible otherwise. The nontrivial solution  of $\Theta(\bvphi)=\bvphi$ disappears following a pitchfork bifurcation~\cite{Strogatz2018} where $\Theta'(\bvphi^*)=1$ and $\Theta(\bvphi^*)=\bvphi^*$, as shown in Fig.~\ref{fig:scheme}(b); $\bvphi^*$ is the discontinuity gap of $\bvphi$. Solving these equations to leading order one finds
\begin{equation}
\bvphi^* \simeq \left\lbrace  
\begin{array}{lll}
\left[\dfrac{(\gamma-1)a_{\gamma-1}}{|a_2|}\right]^{1/(3-\gamma)}  & & 2<\gamma<3\\
&&\\
\left[\dfrac{a_2}{(\gamma-1)|a_{\gamma-1}|}\right]^{1/(\gamma-3)}  & & 3<\gamma<4\\
&&\\
\dfrac{a_2}{3|a_3|}   & & \gamma>4
\end{array}\right..
\label{eq:varphi*}
\end{equation}
For $\gamma>3$, the gap $\bvphi^*$ vanishes for $a_2 = 0$ which leads to $\mus^*=1/2$, the same result of the homogeneous case. For the scale-free regime $2<\gamma<3$, the gap goes to zero when $a_{\gamma-1}=0$. This calculation can be done numerically, as illustrated in Fig.~\ref{fig:betatil}. We have that $\mus^*$ increases non-monotonically from 0  to 1/2 for $\gamma \in (2,3)$ as shown in the phase diagram in parameter's space $\mu_s$ versus $\gamma$ presented in Fig.~\ref{fig:muc}(b). {The scaling of Eq.~\eqref{eq:varphi*} was verified through the numerical solution of Eq.~\eqref{eq:varphi} for large $\kmax$.}

\section{Finite-size scaling in the HMF theory}
\label{sec:numerics}

\begin{figure*}[hbt]
	\centering
	\includegraphics[width=0.315\linewidth]{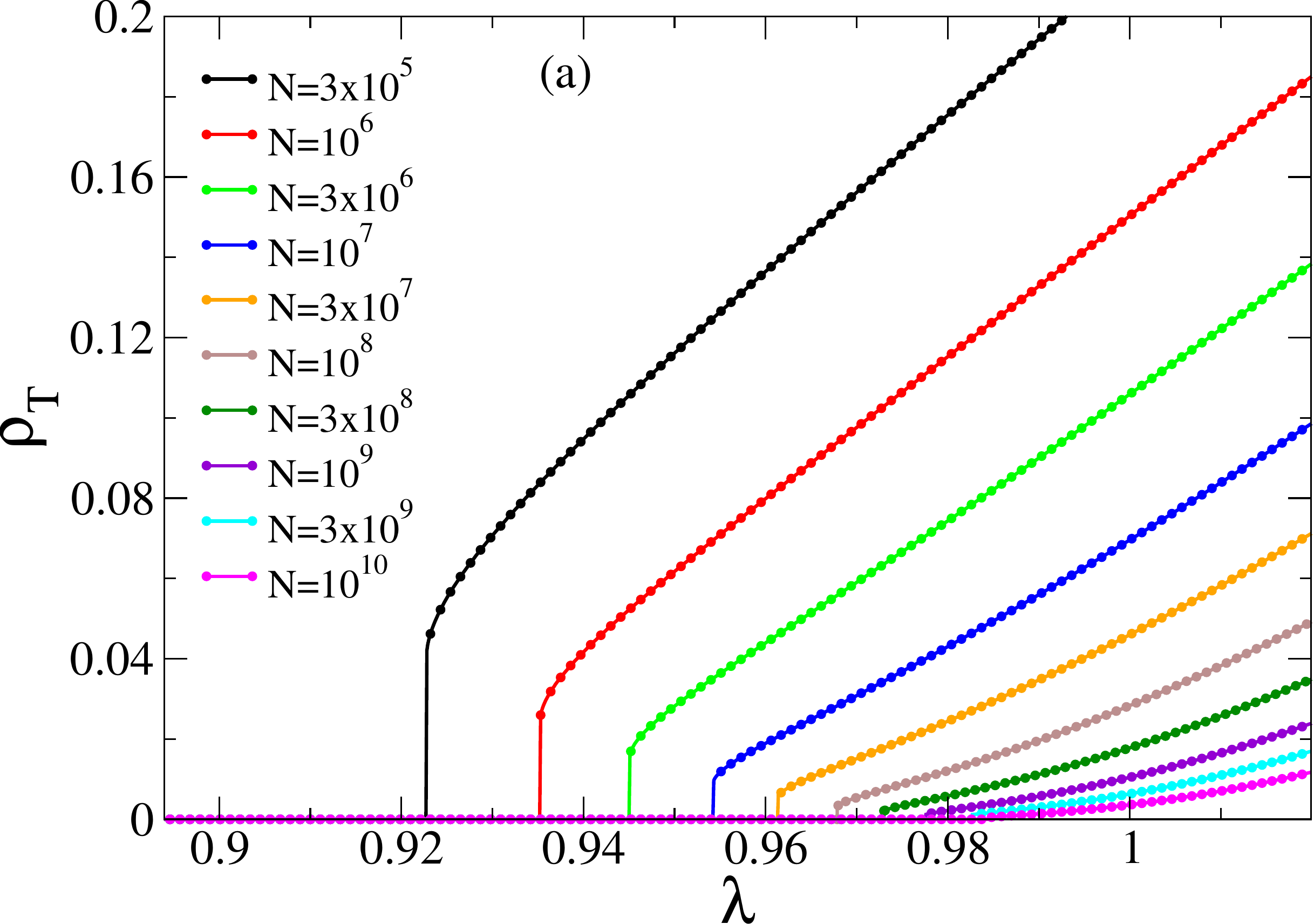}
	\includegraphics[width=0.315\linewidth]{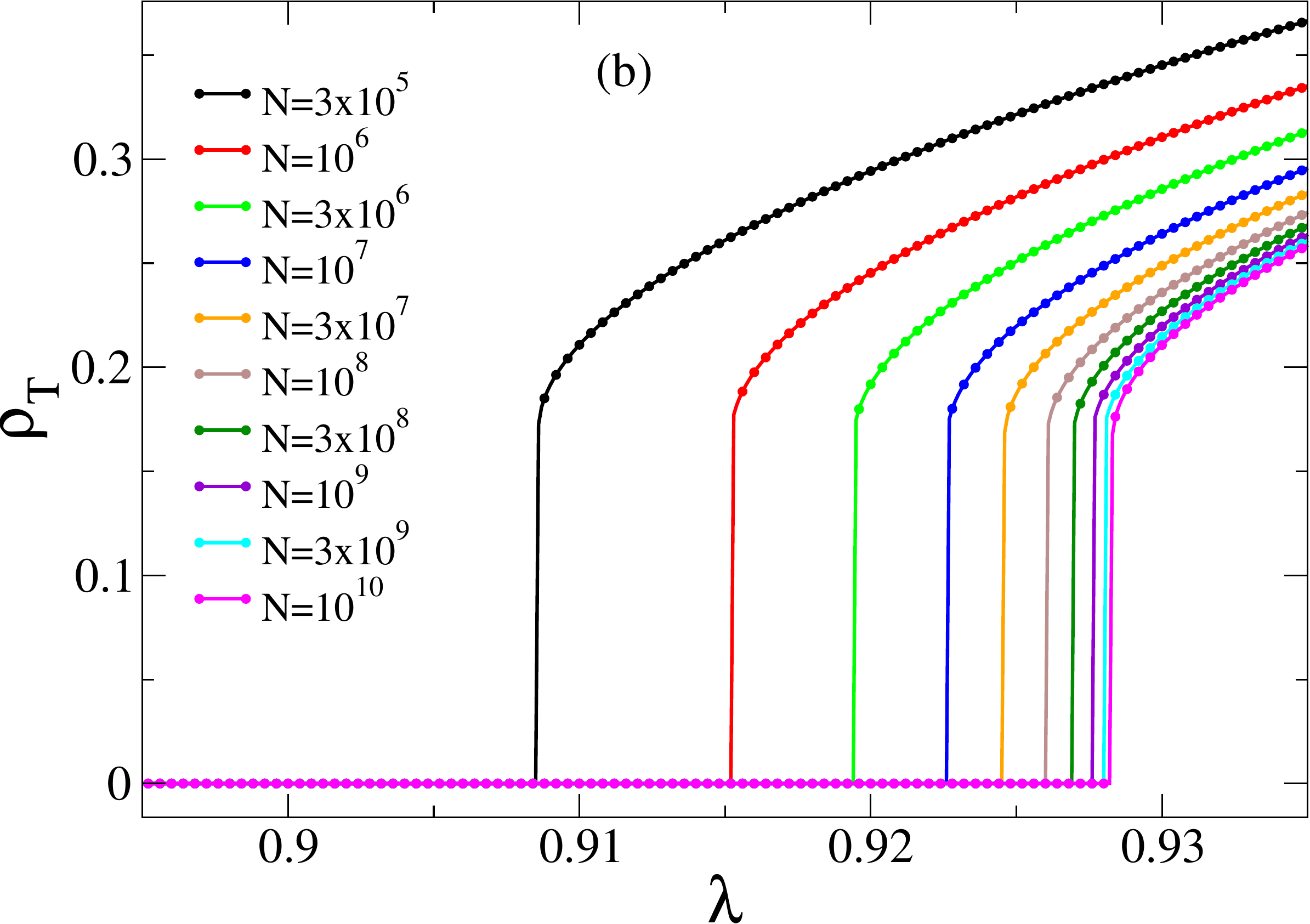}
	\includegraphics[width=0.315\linewidth]{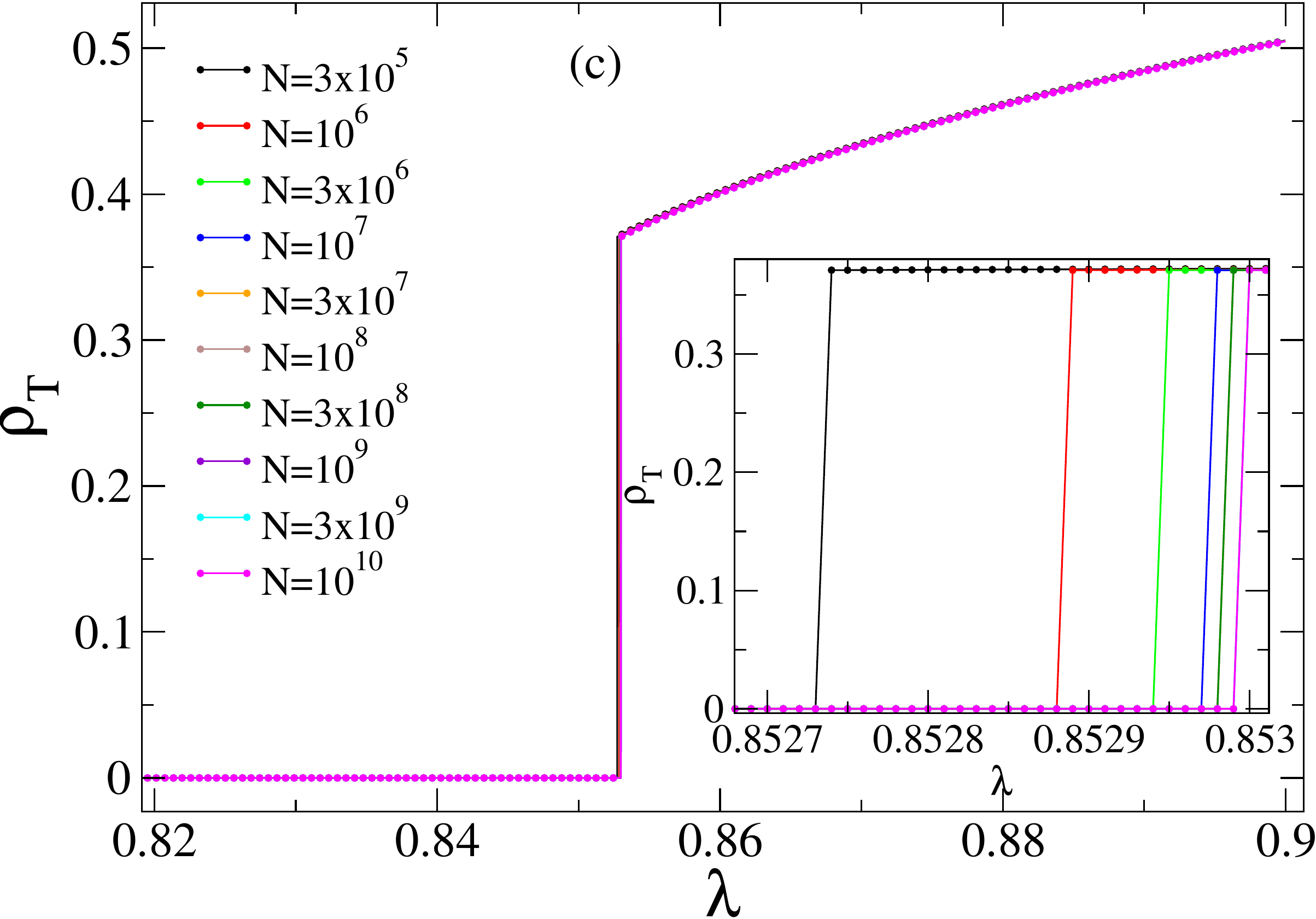}
	\caption{Total prevalence ($\rho_T = 2 \rho + \rab$) as a function of $\lambda$ for $P(k)$ representing synthetic scale-free networks with $\mus = 0.2$ and different sizes. (a) $\gamma = 2.3$, (b) $\gamma = 2.7$. (c) $\gamma = 3.5$}
	\label{fig:intmu02}
\end{figure*}
Stochastic simulations can be performed only in finite size networks. So, it is important to understand the finite size dependence of HMF theory. {So, we} analyze the total prevalence curves corresponding to different  sizes starting from a network {fully occupied by both species} as initial condition. Figure~\ref{fig:intmu02} shows the curves for $\rho_\text{T}$ as function of the infection rate $\lambda$ for $\mus = 0.2$, and different levels of heterogeneity given by $\gamma = 2.3$,  $2.7$ and $3.5$. We observe two finite-size scaling behaviors depending on the degree exponent for a fixed $\mus$. If $\gamma = 2.3$, the transition is discontinuous at an activation threshold $\lambda_\text{c} < 1$ for small sizes. However as $N$ increases, this discontinuity drops towards a continuous  transition in the infinite size limit when $\lambda^-\rightarrow\lambda_\text{c} = 1$ implying a pseudo threshold for finite sizes. On the other hand, for $\gamma =2.7$, the  discontinuity is sustained in the thermodynamic limit, at a threshold converging to  $\lambda^- <  1$. The  convergence can also be seen for $\gamma = 3.5$, being much faster in this case. 

\begin{figure}[hbt]
	\centering
	\includegraphics[width=0.8\linewidth]{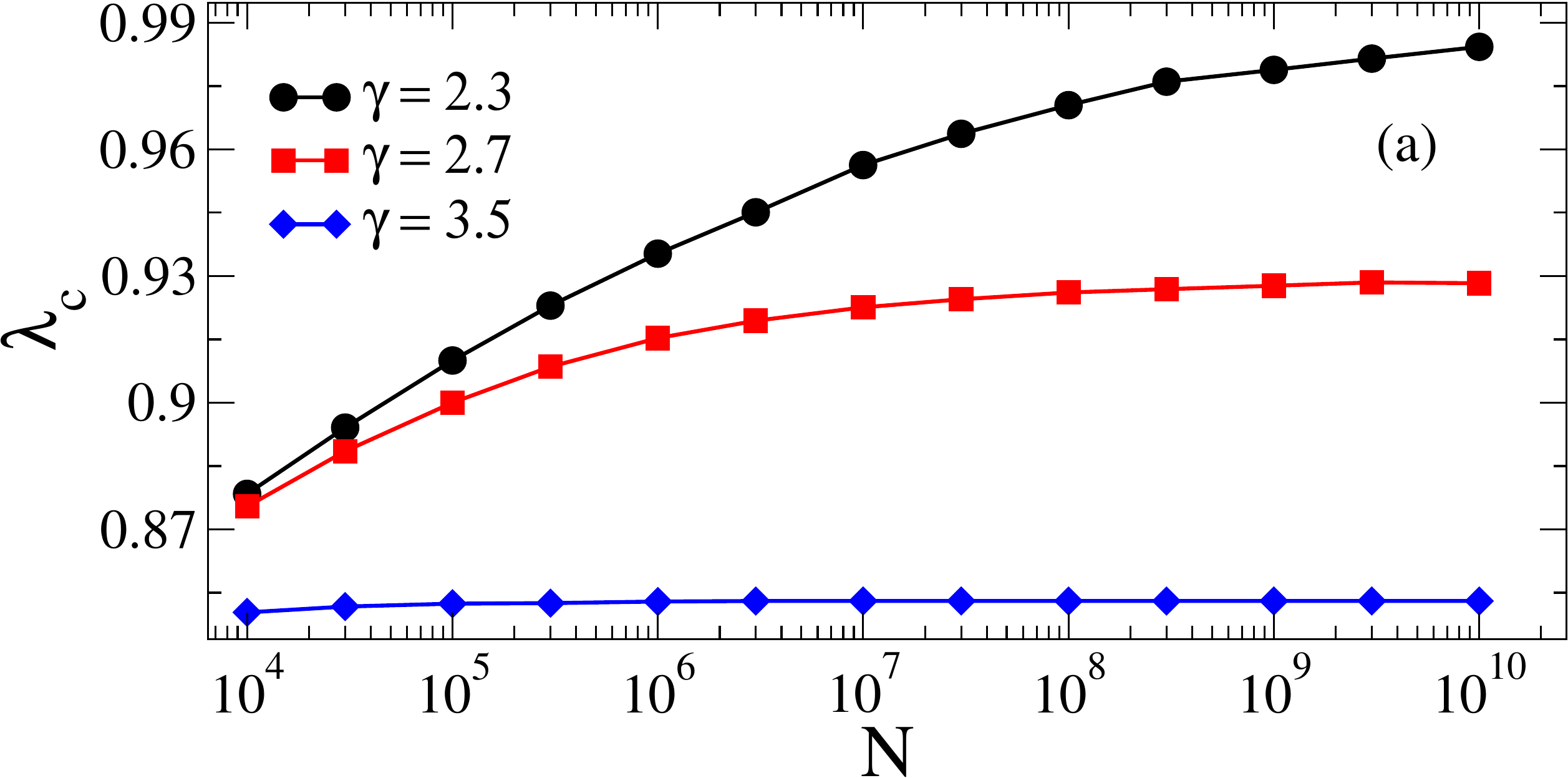}\\
	\includegraphics[width=0.8\linewidth]{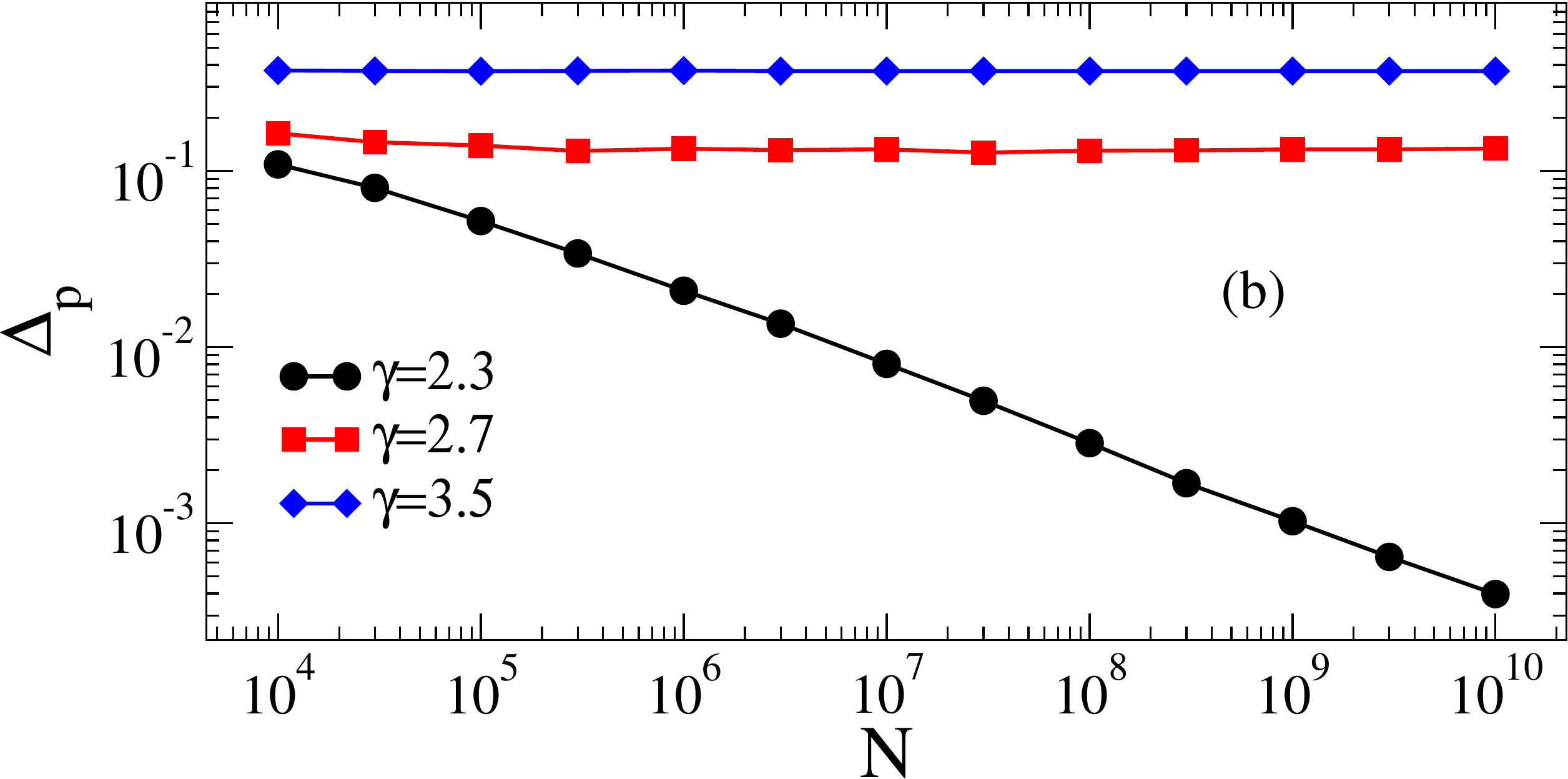}
	\caption{Finite-size scaling for (a) activation threshold and (b) discontinuity gap for 2SCP obtained with the HMF theory for power-law degree distributions with different values of $\gamma$ and $\mus = 0.2$ fixed. Lower and upper degree cutoffs $\kmin=3$ and $\kmax=\sqrt{N}$ we adopted.}
	\label{fig:fss02}
\end{figure}

Figure~\ref{fig:intmu02} presents the finite-size scaling for the activation thresholds  $\lambda_\text{c}$  and the gap discontinuity $\Delta_p$ for curves shown in Fig.~\ref{fig:fss02}. The activation thresholds converge to the  upper spinodal $\lambda^+=1$ for $\gamma=2.3$ while the convergence to the lower spinodals are reported for $\gamma=2.7$ and 3.5. The discontinuity gap remains finite for $\gamma = 2.7$ and $3.5$, but  decay as a power-law $\Delta_\text{p} \sim N^{-0.43}$, corroborating the transition continuity as $N \to \infty$.

\begin{figure}[hbt]
	\centering
	\includegraphics[width=0.8\linewidth]{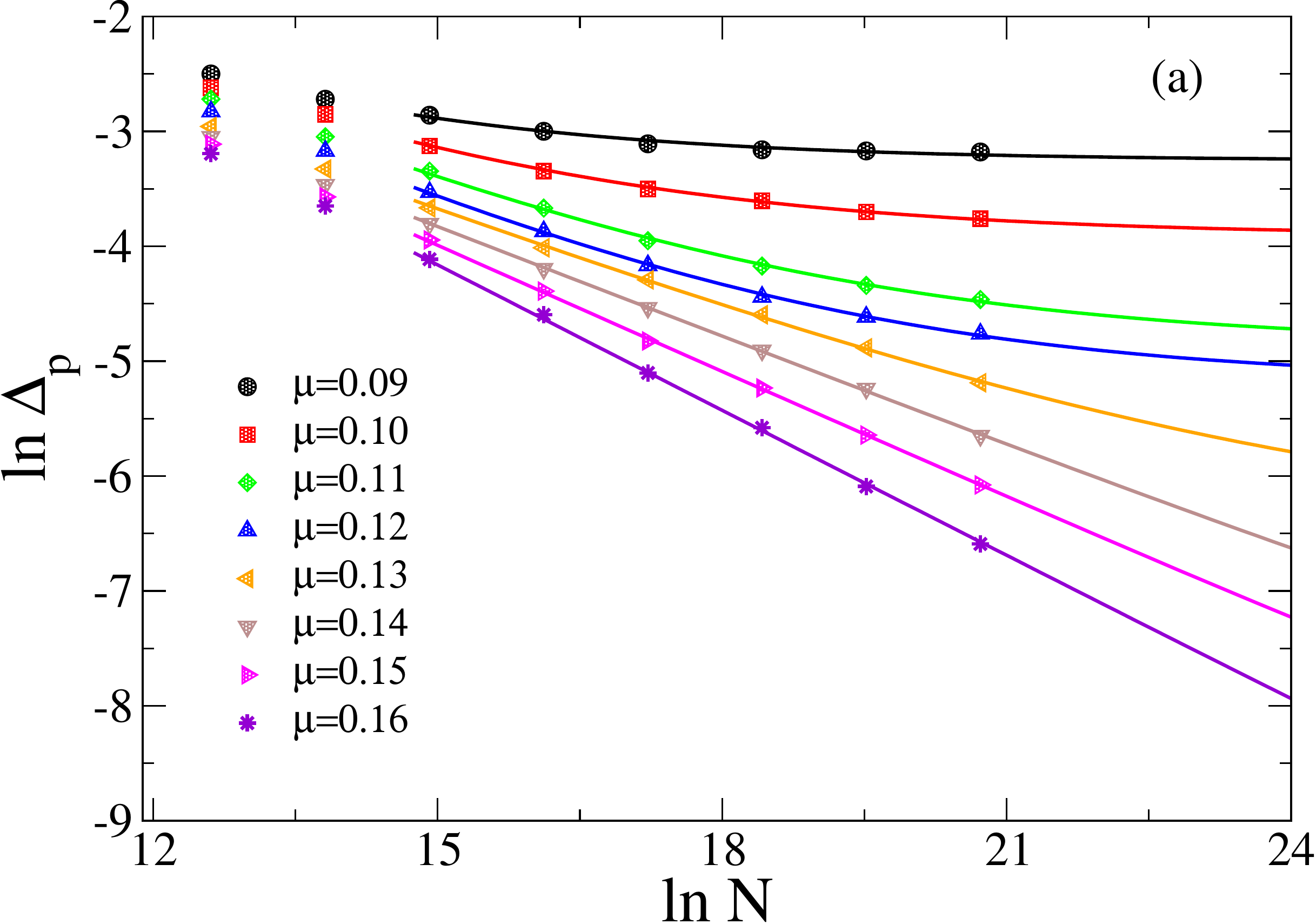}
	\includegraphics[width=0.8\linewidth]{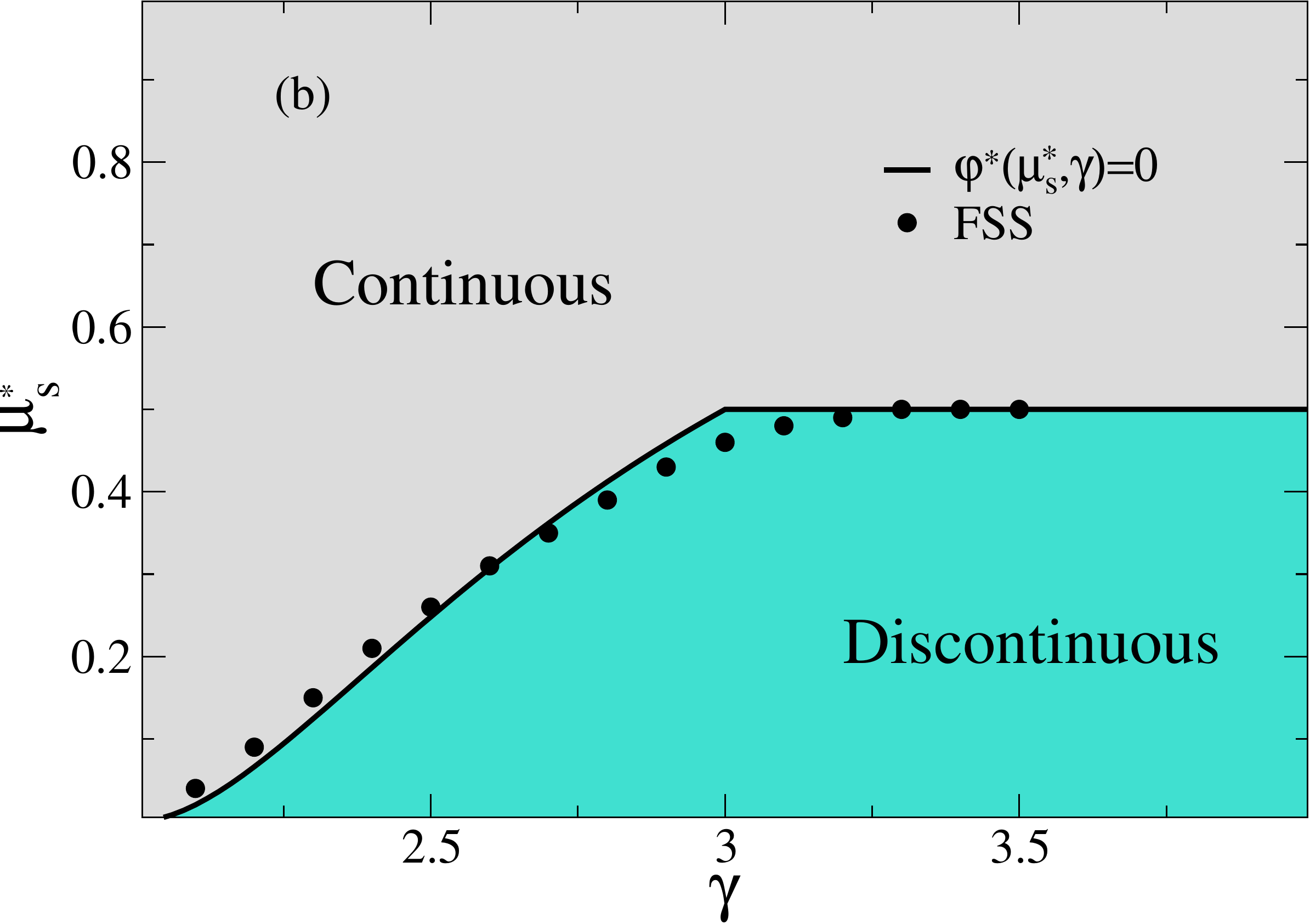}
	\caption{(a) Finite-size scaling (FSS) of discontinuity gap ($\Delta_\text{p}$) for 2SCP obtained with the HMF theory for power-law degree distributions with  $\gamma = 2.3$ and different values of $\mus$. {Symbols are numerical data obtained from integration of HMF equations and solid lines non-linear regression to perform the FSS; see main text.}  (b) Symbiosis coupling $\mus^*$ separating  the discontinuous and continuous transitions  for different values of the degree exponent, obtained {theoretically (solid curve) and} {with the} {FSS} of $\Delta_\text{p}$ versus $N$. }
	\label{fig:muc}
\end{figure}

{We} determine {numerically} the value $\mus^*$ of the symbiotic coupling that separates the continuous from discontinuous transitions, analyzing different levels of heterogeneity. Figure \ref{fig:muc}(a) shows the finite-size scaling of $\Delta_p$ for $\gamma = 2.3$ and different values of $\mus$. While curves for lower values of $\mus$ tend to a finite value of $\Delta_p$ (bends upwards), it keeps decaying towards 0,  within the accuracy of our numerical solution, for higher values. {To estimate the value of $\mus^*$, we assume a monotonic approximation of the asymptotic limit using an scaling in form $\Delta_\text{p} (N) = \Delta_\text{p} (\infty)+c N^{-b}$ or, more precisely, $z=z_0+\ln(1+ce^{-b w})$ with $z=\ln(\Delta_\text{p})$ and $w=\ln(N)$ in double logarithmic form to reckon correctly the scaling behavior.}  By collecting the values of $\mus^*$ for different $\gamma$, we obtain the phase diagram in the space parameter $\mus$ versus $\gamma$ presented in Fig.\ref{fig:muc}(b), in which the {theoretical} curve that separates the continuous from discontinuous transitions is presented. {We observe a very good match between finite-size scaling and the theoretical prediction, with more significant differences for $\gamma$ close to 2 or 3. Indeed, this same effect was reported for the ordinary contact processes on networks, presenting stronger finite-size corrections in these ranges of $\gamma$~\cite{Ferreira2011a,Mata2014}. This agreement  qualifies the method to the analysis of stochastic simulations where no exact expression is available.} Figure~\ref{fig:muc}(b) also corroborates the shrinking of the bistability region in the phase parameter $\lambda$ versus $\mus$ as the heterogeneity is increased towards $\gamma=2$ {and the independence of $\mus^*$ on the degree exponent for $\gamma>3$. Note that, according to Eq.~\eqref{eq:varphi*}, heterogeneity alters the critical dependence for $3<\gamma<4$ where the gap  $\bvphi^*$ goes to zero following a scaling that depend on $\gamma$, while the homogeneous mean-field behavior is recovered only for $\gamma>4$.}

\section{Quasistationary simulations for 2SCP on complex networks}
\label{sec:simu}

\subsection{Methods}

In order to validate the predictions of the HMF theory, we performed simulations of 2SCP on annealed and quenched networks with power-law degree distributions. Connections in annealed networks are probabilistic, such that a node $i$ can be connected  to any other node $j$  with probability proportional to the product $k_ik_j$ in a given time step~\cite{Boguna2009}. Due to the constant rewiring, annealed networks are substrates in which HMF theories are expected to be exact in the thermodynamic limit and can be used to corroborate the correctness of the HMF equations~\cite{Ferreira2011a,Boguna2009}. Annealed networks are easily implemented attributing the desired degree sequence $\lbrace k_1, \ldots, k_N\rbrace$  for the nodes of the network. When a neighbor has to be chosen, a vertex is randomly selected with probability  proportional to its degree~\cite{Boguna2009}. The uncorrelated  configuration model (UCM)~\cite{Catanzaro2005}  was used to simulate quenched networks.  An upper cutoff $\kmax=\sqrt{N}$ guarantees absence of degree correlations and was used to allow a comparison with annealed and HMF results. 

We performed stochastic simulations using an optimized Gillespie algorithm detailed in Ref.~\cite{deOliveira2019}.To deal with the intrinsic difficulties of absorbing states in finite-size systems~\cite{Marro2005}, we adopt the slightly modified QS simulations~\cite{Costa2021,DeOliveira2005}, constraining the averages to configurations in which none species is extinct~\cite{deOliveira2019}. A list with $M_\text{conf}$ configurations visited along the dynamics, in which both species are active, is constructed. This list is constantly updated by replacing one of its elements with the current configuration with probability $q$ per unit of time. One configuration of this list is randomly chosen to replace the system's state whenever one of the species is extinct. The QS quantities are computed after a relaxation time $t_\text{rlx}$ during an average time $t_\text{av}$. The QS method has been successively applied to diverse dynamical processes with absorbing states on complex networks~\cite{Ferreira2011, Mata2014, deOliveira2019, DeArruda2015a, Sander2016}. In the present work we used $M=100$, $q=10^{-2}$, $t_\text{rlx}=10^6$, and $t_\text{av}=10^7$.

\begin{figure}[hbt]
	\centering
	\includegraphics[width=0.8\linewidth]{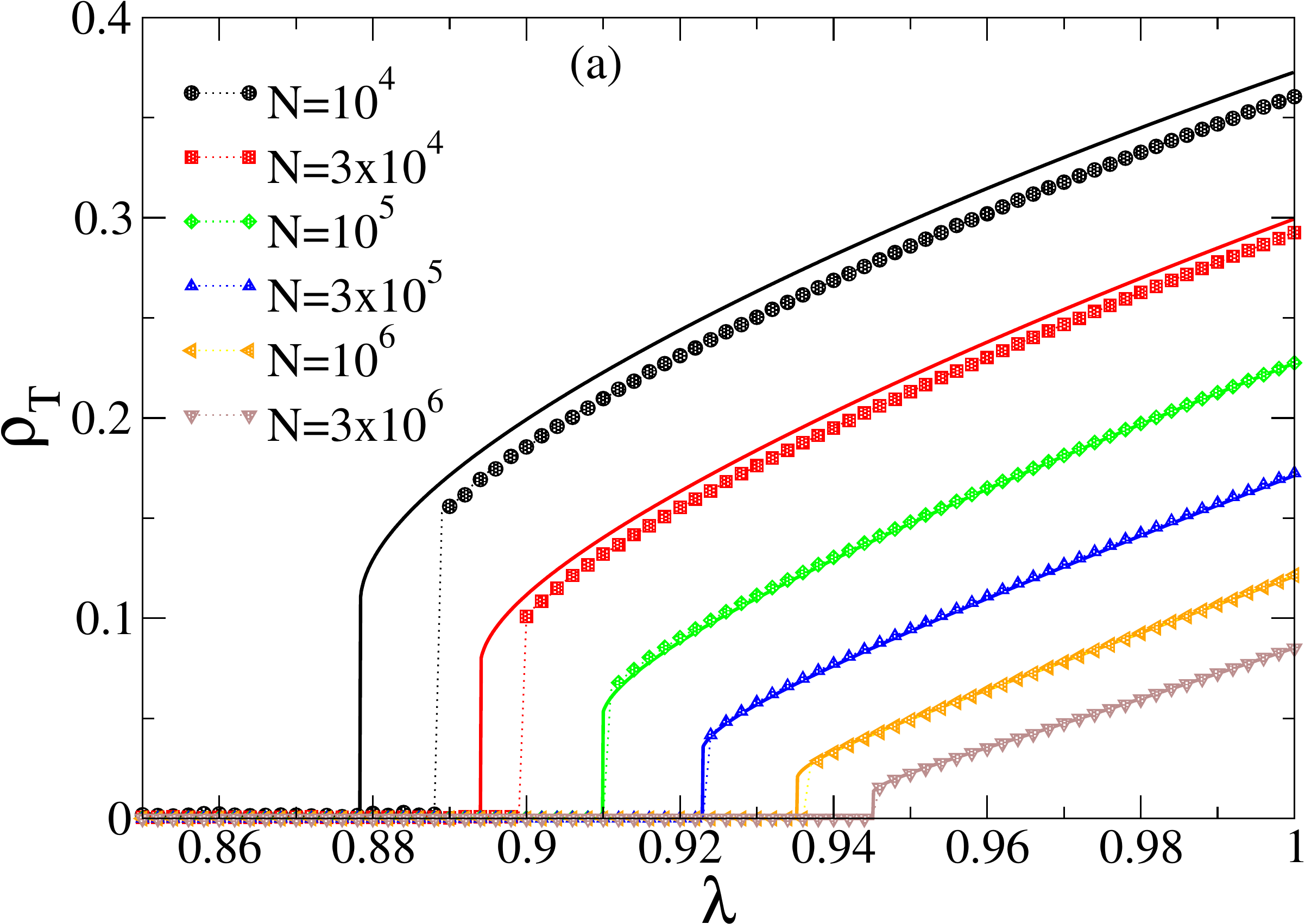}
	\includegraphics[width=0.8\linewidth]{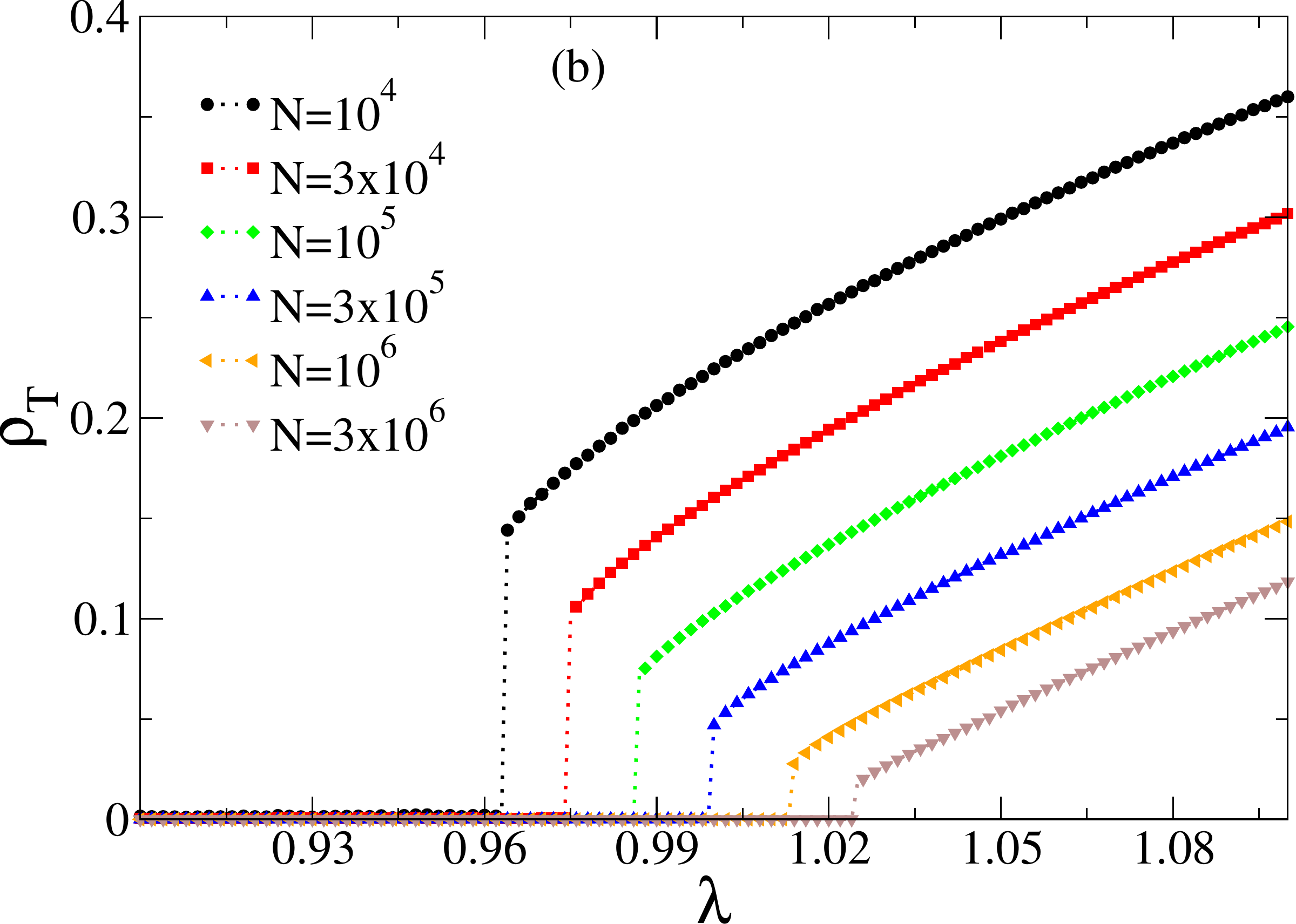}
	\caption{(a)Total prevalence curves obtained with QS simulations of the  2SCP model on (a) annealed and (b) quenched networks  with degree exponent $\gamma=2.3$ and different sizes (symbols) in comparison with HMF equations integration of the same degree distribution $P(k)$ (solid lines). The lower and upper degree cutoff are $\kmin=3$ and $\kmax=\sqrt{N}$ and the symbiosis parameter is $\mus=0.2$. Doted lines are guides to  the eyes to visualize the discontinuity in simulation curves.}
	\label{fig:simuann}
\end{figure}

\subsection{Results}

Sizes attainable in stochastic simulations are much smaller than in the solution of the HMF equation~\eqref{eq:varphi} that involve  $n\sim \kmax$ variables while in the simulations the number of agents scale as $N=(\kmax)^2$. Additionally, the RAM memory demanded can be orders of magnitude larger. The comparison between total prevalence obtained with HMF theory and QS simulations on annealed networks of different sizes with degree exponent $\gamma=2.3$  is shown in Fig.~\ref{fig:simuann}(a). One can see that the simulations converge to the theoretical predictions as $N$ increases while the discrepancies for smaller sizes are due to the absence of stochasticity in the theory, which become negligible in the infinite-size limit. {Higher values of $\gamma$ lead to even better agreement of QS simulation with HMF theory and the results are physically similar to the presented case $\gamma=2.3$.} This agreement corroborates the correctness of the HMF analysis.
\begin{figure}[hbt]
	\centering
	\includegraphics[width=0.8\linewidth]{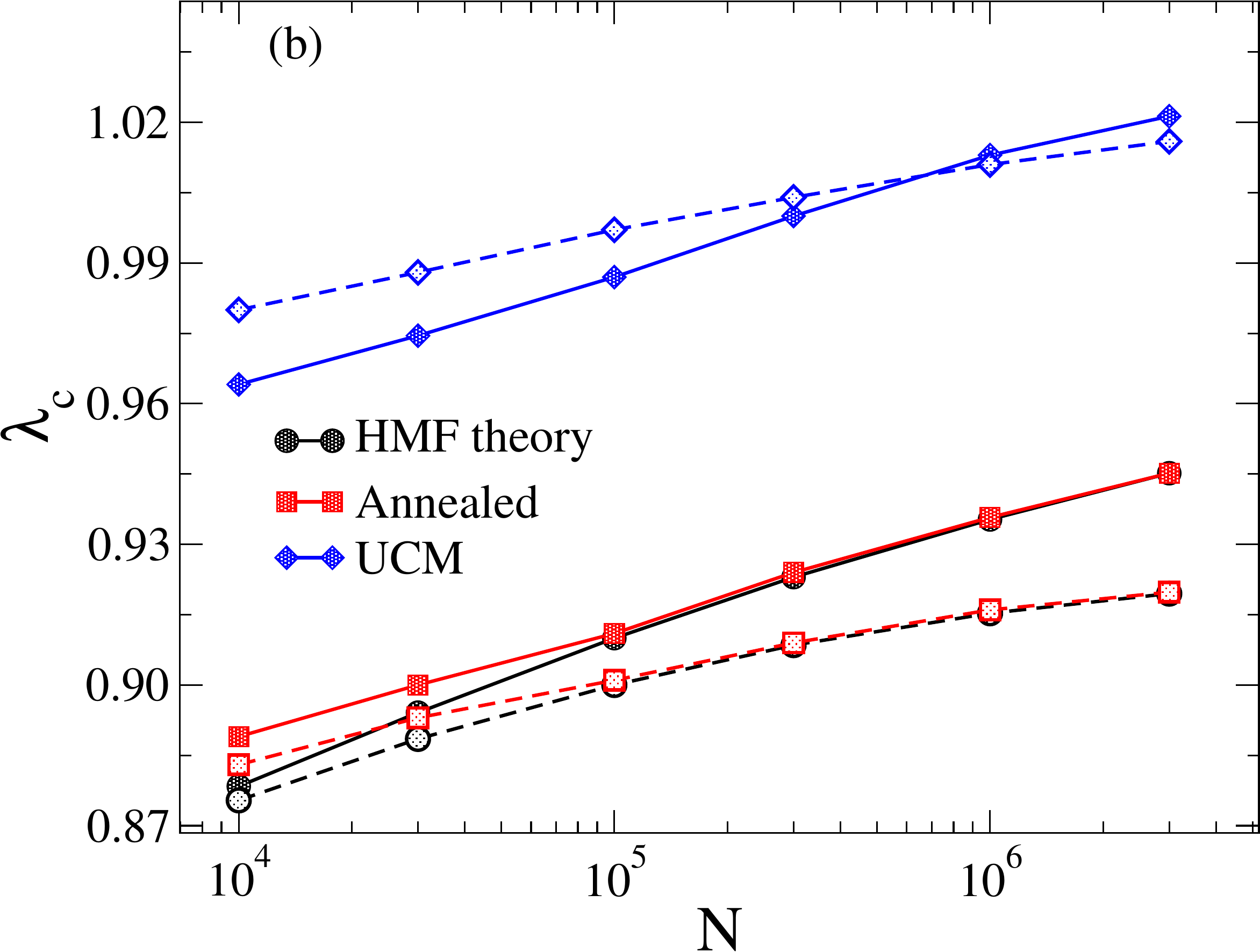}
	\includegraphics[width=0.8\linewidth]{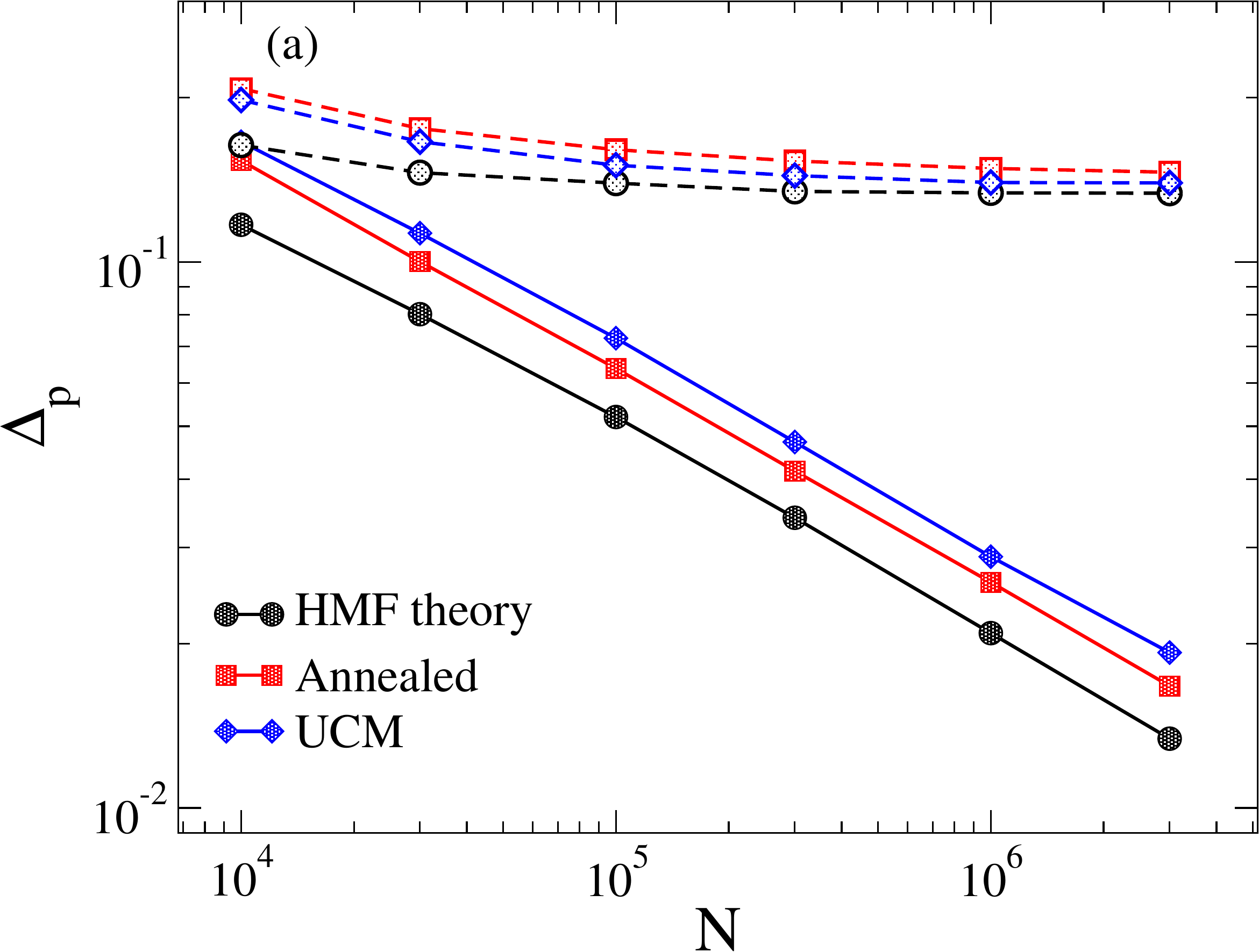}
	\caption{Comparison of the finite-size scaling of the (a) gap discontinuity $\Delta_p$ and the (b) activation threshold $\lambda_\text{c}$ obtained in HMF theory and simulations on quenched and annealed networks with $\gamma = 2.3$ (solid lines) and $2.7$ (dashed lines). The color sequence is black for HMF theory, red for annealed and blue for UCM networks. The lower and upper cutoffs are $\kmin=3$ and $\kmax=\sqrt{N}$, respectively, and the symbiosis parameter is $\mus=0.2$.}
	\label{fig:fsscomp}
\end{figure}
Simulations of 2SCP on UCM quenched networks with the same parameters of the annealed case are presented in Fig.\ref{fig:simuann}(b).  While a similar qualitative behavior is observed, with the reduction of discontinuity gap as $N$ increases, there are expected quantitative differences  such as the shift of the activation threshold  in the lower spinodal towards higher values. Indeed, the dynamical correlations existing in quenched networks play the important role of raising the activation threshold as observed for the single species CP~\cite{Ferreira2011,Mata2014} and 2SCP \cite{deOliveira2019}.

Figure \ref{fig:fsscomp}(a) shows the finite-size scaling for both HMF theory and simulations on power-law networks. The activation threshold for annealed networks converges quickly to the HMF theory while the UCM one deviates from the latter. The discontinuity gap presents the same behavior for all three approaches and both degree exponents which were investigated. The small differences between annealed networks and HMF theory can be attributed to the limited accuracy in determining the activation threshold in simulations for which a small uncertainty leads to imprecision in the critical quantities.

\section{Summary and discussions}
\label{sec:conclu}

Coexistence of dynamical processes on the top of complex networks is a breakthrough issue that has been investigated in several applied contexts, which demand more complex models and substrates.  So, due to its simplicity, the two species contact process can be used as a benchmark model to investigate the nature of discontinuous or continuous transition in coexisting dynamics. The existence of continuous and discontinuous transitions in the 2SCP has been already theorized in simpler mean-field theories, such as the homogeneous one~\cite{deOliveira2012,deOliveira2014}, in which the discontinuity threshold is fixed and independent of the substrate. Even for more complicated theories, such as the pairwise mean-field theory~\cite{deOliveira2019}, this threshold is independent of the network heterogeneity. The role of heterogeneity of the contact structure has not been analyzed from the theoretical point of view. In the present work, we tackle this problem by investigating a heterogeneous mean-field theory for the 2SCP  with a set of equations that take into account the degree  distribution. 

The HMF theory predicts that increased heterogeneity leads to the reduction of the regions of bistability in the phase diagram $\lambda$ versus $\mus$, where absorbing and active states are locally stable. In terms of network parameters, the bistability region shrinks as the degree exponent $\gamma$ goes to the lower bound $\gamma=2$. Weakening of bistability can be attributed to the lowered probability to produce  doubly occupied nodes among the neighbors of a hub, which are the most active elements of the network. We performed a careful finite-size analysis and observed complex behaviors. Depending on the strength of the symbiotic coupling, a discontinuous transition observed at a finite size gradually converges to a continuous one in the thermodynamic limit, consisting in a pseudo discontinuous transition at finite sizes. Our analytical results are backed up by extensive stochastic simulations on both annealed and quenched networks.

We expect that our results will stimulate further fundamental investigations of the interplay between heterogeneity and bistability in nonequilibrium absorbing-state phase transitions and can help to understand basic phenomena in applied modeling of coexisting dynamics such as rumors and contagious diseases~\cite{Bedson2021}, interacting diseases~\cite{Newman2013,Sanz2014}, and ecological symbiosis~\cite{Iwata2011}. {Natural sequence of the current work is to extend the theory to networks with degree correlations observed in many real networks~\cite{barabasi2016network}.}
  
\appendix

\begin{widetext}
\section{Continuous approximation for Eq.~\eqref{eq:varphi}} 
\label{app:HMF}
Starting with Eq~\eqref{eq:varphi_cont}, we complete squares in the denominator and takes a binomial series expansion to find
\begin{equation}
\Theta(\bvphi) = \int_{\kmin}^{\infty} \dfrac{\beta_k\bvphi(\sqrt{\mus}+\bvphi\beta_k/\sqrt{\mus})}{\sqrt{\mus}(1+\bvphi\beta_k/\sqrt{\mus})^2}\sum_{n=0}^{\infty}  \left[\frac{2(1-\sqrt{\mus})\bvphi \beta_k/\sqrt{\mus}}{(1+\bvphi\beta_k/\sqrt{\mus})^2}\right]^nP(k) dk.
\end{equation}
The sum is converging since $\zeta=2\bvphi(\beta_k/\sqrt{\mus})/(1+\bvphi\beta_k/\sqrt{\mus})^2\le 1/2$ was used in series expansion of $(1-\zeta)^{-1}$. Now, considering a normalized power-law distribution $P(k)=Ak^{-\gamma}$ with $A=(\gamma-1)\kmin^{\gamma-1}$, the change of variable $x= \lambda k/(\sqrt{\mus}\av{k})$ leads to 
\begin{equation}
\Theta(\bvphi) = \alpha \sum_{n=0}^{\infty}[2(1-\sqrt{\mus})]^{n} \bvphi^{n+1}
\left[\sqrt{\mus}\int_{x_0}^{\infty} \frac{x^{-\gamma+n+1}}{(1+\bvphi x)^{2n+2}}+\bvphi \frac{x^{-\gamma+n+2}}{(1+\bvphi x)^{2n+2}} \right],
\label{eq:appTheta1}
\end{equation}
where $x_0=\lambda\kmin/\sqrt{\mus\av{k}}$ and $\alpha = (\gamma-1)\left({\kmin\lambda}/{\sqrt{\mus}\av{k}}\right)^{\gamma-1}$. The integrals in Eq.~\eqref{eq:appTheta1} can be expressed in terms of Gauss hypergeometric functions $F(a,b,c;x)$~\cite{zwillinger2007table}, to obtain
\begin{align}
\Theta(\bvphi) = & \alpha x_0^{-\gamma+1} \sum_{n=0}^{\infty}[2(1-\sqrt{\mus})]^{n} (\bvphi x_0)^{-(n+1)} 
\left[ \frac{\sqrt{\mus}}{n+\gamma} F\left(2n\!+\!2,n+\!\gamma,n+\gamma+1;\dfrac{-1}{\bvphi x_0}\right)\right. \nonumber \\
+ & \left. \frac{\bvphi x_0}{n+\gamma-1} F\left(2n+2,n+\gamma-1,n+\gamma;\frac{-1}{\bvphi x_0}\right)
\right].
\end{align}
Now we take the asymptotic expansion for $F(a,b,c;z)$ for $z\rightarrow\infty$ up order $\bvphi^3$ to obtain
\begin{equation}
\Theta(\bvphi)=\lambda\bvphi+ a_{\gamma-1} \bvphi^{\gamma-1}+\frac{(\gamma-2)^2(1-2\mus)}{(\gamma-3)(\gamma-1)\mus}\lambda^2\bvphi^2
+\frac{(\gamma-2)^3(7-8\sqrt{\mus})}{(4-\gamma)(\gamma-1)^2\mus}\lambda^3 \bvphi^3+\cdots,
\label{eq:Theta_series}
\end{equation}
where
\begin{equation}
a_{\gamma-1} =  (\gamma-1)
\left(\frac{\lambda\kmin}{\av{k}\sqrt{\mus}}\right)^{\gamma-1} \tilde{\beta}
\label{eq:a_gamma_m_1}
\end{equation}
and
\begin{equation}
\tilde{\beta} = \sum_{n=0}^{\infty}[2(1\!-\!\sqrt{\mus})]^n\left[\sqrt{\mus}\!+\!
\frac{n-\gamma+2}{n+\gamma-1} \right] \frac{\Gamma(n-\gamma+2)\Gamma(\gamma+n)}{\Gamma(2n+2)},
\label{eq:betatil}
\end{equation}
where $\Gamma(x)$ is the Gamma function~\cite{zwillinger2007table}.
\end{widetext}

\begin{acknowledgments}
{SCF and GSC thanks the support by the Brazilian agencies
\textit{Conselho Nacional de Desenvolvimento Científico e Tecnológico}- CNPq
(Grants no. 430768/2018-4 and 311183/2019-0) and \textit{Fundação de Amparo à
Pesquisa do Estado de Minas Gerais} - FAPEMIG (Grant no. APQ-02393-18).} This
study was financed in part by the \textit{Coordenação de Aperfeiçoamento de
Pessoal de Nível Superior} (CAPES) - Brasil  - Finance Code 001.
MMO thanks the support of CNPq (Grant no. 304909/2018-1).
\end{acknowledgments}

%\bibliography{cpsimb}

\begin{thebibliography}{44}%
	\makeatletter
	\providecommand \@ifxundefined [1]{%
		\@ifx{#1\undefined}
	}%
	\providecommand \@ifnum [1]{%
		\ifnum #1\expandafter \@firstoftwo
		\else \expandafter \@secondoftwo
		\fi
	}%
	\providecommand \@ifx [1]{%
		\ifx #1\expandafter \@firstoftwo
		\else \expandafter \@secondoftwo
		\fi
	}%
	\providecommand \natexlab [1]{#1}%
	\providecommand \enquote  [1]{``#1''}%
	\providecommand \bibnamefont  [1]{#1}%
	\providecommand \bibfnamefont [1]{#1}%
	\providecommand \citenamefont [1]{#1}%
	\providecommand \href@noop [0]{\@secondoftwo}%
	\providecommand \href [0]{\begingroup \@sanitize@url \@href}%
	\providecommand \@href[1]{\@@startlink{#1}\@@href}%
	\providecommand \@@href[1]{\endgroup#1\@@endlink}%
	\providecommand \@sanitize@url [0]{\catcode `\\12\catcode `\$12\catcode
		`\&12\catcode `\#12\catcode `\^12\catcode `\_12\catcode `\%12\relax}%
	\providecommand \@@startlink[1]{}%
	\providecommand \@@endlink[0]{}%
	\providecommand \url  [0]{\begingroup\@sanitize@url \@url }%
	\providecommand \@url [1]{\endgroup\@href {#1}{\urlprefix }}%
	\providecommand \urlprefix  [0]{URL }%
	\providecommand \Eprint [0]{\href }%
	\providecommand \doibase [0]{https://doi.org/}%
	\providecommand \selectlanguage [0]{\@gobble}%
	\providecommand \bibinfo  [0]{\@secondoftwo}%
	\providecommand \bibfield  [0]{\@secondoftwo}%
	\providecommand \translation [1]{[#1]}%
	\providecommand \BibitemOpen [0]{}%
	\providecommand \bibitemStop [0]{}%
	\providecommand \bibitemNoStop [0]{.\EOS\space}%
	\providecommand \EOS [0]{\spacefactor3000\relax}%
	\providecommand \BibitemShut  [1]{\csname bibitem#1\endcsname}%
	\let\auto@bib@innerbib\@empty
	%</preamble>
	\bibitem [{\citenamefont {Wang}\ \emph {et~al.}(2019)\citenamefont {Wang},
		\citenamefont {Liu}, \citenamefont {Liang}, \citenamefont {Hu},\ and\
		\citenamefont {Zhou}}]{Wang2019}%
	\BibitemOpen
	\bibfield  {author} {\bibinfo {author} {\bibfnamefont {W.}~\bibnamefont
			{Wang}}, \bibinfo {author} {\bibfnamefont {Q.-H.}\ \bibnamefont {Liu}},
		\bibinfo {author} {\bibfnamefont {J.}~\bibnamefont {Liang}}, \bibinfo
		{author} {\bibfnamefont {Y.}~\bibnamefont {Hu}},\ and\ \bibinfo {author}
		{\bibfnamefont {T.}~\bibnamefont {Zhou}},\ }\bibfield  {title} {\bibinfo
		{title} {{Coevolution spreading in complex networks}},\ }\href
	{https://linkinghub.elsevier.com/retrieve/pii/S0370157319302583} {\bibfield
		{journal} {\bibinfo  {journal} {Phys. Rep.}\ }\textbf {\bibinfo {volume}
			{820}},\ \bibinfo {pages} {1} (\bibinfo {year} {2019})}\BibitemShut {NoStop}%
	\bibitem [{\citenamefont {Iwata}\ \emph {et~al.}(2011)\citenamefont {Iwata},
		\citenamefont {Kobayashi}, \citenamefont {Higa}, \citenamefont {Yoshimura},\
		and\ \citenamefont {Tainaka}}]{Iwata2011}%
	\BibitemOpen
	\bibfield  {author} {\bibinfo {author} {\bibfnamefont {S.}~\bibnamefont
			{Iwata}}, \bibinfo {author} {\bibfnamefont {K.}~\bibnamefont {Kobayashi}},
		\bibinfo {author} {\bibfnamefont {S.}~\bibnamefont {Higa}}, \bibinfo {author}
		{\bibfnamefont {J.}~\bibnamefont {Yoshimura}},\ and\ \bibinfo {author}
		{\bibfnamefont {K.-i.}\ \bibnamefont {Tainaka}},\ }\bibfield  {title}
	{\bibinfo {title} {{A simple population theory for mutualism by the use of
				lattice gas model}},\ }\href
	{https://linkinghub.elsevier.com/retrieve/pii/S0304380011002171} {\bibfield
		{journal} {\bibinfo  {journal} {Ecol. Modell.}\ }\textbf {\bibinfo {volume}
			{222}},\ \bibinfo {pages} {2042} (\bibinfo {year} {2011})}\BibitemShut
	{NoStop}%
	\bibitem [{\citenamefont {Dobramysl}\ and\ \citenamefont
		{T\"auber}(2013)}]{Ulrich2015}%
	\BibitemOpen
	\bibfield  {author} {\bibinfo {author} {\bibfnamefont {U.}~\bibnamefont
			{Dobramysl}}\ and\ \bibinfo {author} {\bibfnamefont {U.~C.}\ \bibnamefont
			{T\"auber}},\ }\bibfield  {title} {\bibinfo {title} {Environmental versus
			demographic variability in two-species predator-prey models},\ }\href
	{https://link.aps.org/doi/10.1103/PhysRevLett.110.048105} {\bibfield
		{journal} {\bibinfo  {journal} {Phys. Rev. Lett.}\ }\textbf {\bibinfo
			{volume} {110}},\ \bibinfo {pages} {048105} (\bibinfo {year}
		{2013})}\BibitemShut {NoStop}%
	\bibitem [{\citenamefont {Newman}(2005)}]{Newman2005}%
	\BibitemOpen
	\bibfield  {author} {\bibinfo {author} {\bibfnamefont {M.~E.~J.}\
			\bibnamefont {Newman}},\ }\bibfield  {title} {\bibinfo {title} {{Threshold
				Effects for Two Pathogens Spreading on a Network}},\ }\href
	{https://link.aps.org/doi/10.1103/PhysRevLett.95.108701} {\bibfield
		{journal} {\bibinfo  {journal} {Phys. Rev. Lett.}\ }\textbf {\bibinfo
			{volume} {95}},\ \bibinfo {pages} {108701} (\bibinfo {year}
		{2005})}\BibitemShut {NoStop}%
	\bibitem [{\citenamefont {Newman}\ and\ \citenamefont
		{Ferrario}(2013)}]{Newman2013}%
	\BibitemOpen
	\bibfield  {author} {\bibinfo {author} {\bibfnamefont {M.~E.~J.}\
			\bibnamefont {Newman}}\ and\ \bibinfo {author} {\bibfnamefont {C.~R.}\
			\bibnamefont {Ferrario}},\ }\bibfield  {title} {\bibinfo {title}
		{{Interacting Epidemics and Coinfection on Contact Networks}},\ }\href
	{https://dx.plos.org/10.1371/journal.pone.0071321} {\bibfield  {journal}
		{\bibinfo  {journal} {PLoS One}\ }\textbf {\bibinfo {volume} {8}},\ \bibinfo
		{pages} {e71321} (\bibinfo {year} {2013})}\BibitemShut {NoStop}%
	\bibitem [{\citenamefont {Bianconi}\ \emph {et~al.}(2021)\citenamefont
		{Bianconi}, \citenamefont {Sun}, \citenamefont {Rapisardi},\ and\
		\citenamefont {Arenas}}]{Bianconi2021}%
	\BibitemOpen
	\bibfield  {author} {\bibinfo {author} {\bibfnamefont {G.}~\bibnamefont
			{Bianconi}}, \bibinfo {author} {\bibfnamefont {H.}~\bibnamefont {Sun}},
		\bibinfo {author} {\bibfnamefont {G.}~\bibnamefont {Rapisardi}},\ and\
		\bibinfo {author} {\bibfnamefont {A.}~\bibnamefont {Arenas}},\ }\bibfield
	{title} {\bibinfo {title} {Message-passing approach to epidemic tracing and
			mitigation with apps},\ }\href
	{https://doi.org/10.1103/PhysRevResearch.3.L012014} {\bibfield  {journal}
		{\bibinfo  {journal} {Phys. Rev. Research}\ }\textbf {\bibinfo {volume}
			{3}},\ \bibinfo {pages} {L012014} (\bibinfo {year} {2021})}\BibitemShut
	{NoStop}%
	\bibitem [{\citenamefont {Granell}\ \emph {et~al.}(2013)\citenamefont
		{Granell}, \citenamefont {G\'omez},\ and\ \citenamefont
		{Arenas}}]{Granell2013}%
	\BibitemOpen
	\bibfield  {author} {\bibinfo {author} {\bibfnamefont {C.}~\bibnamefont
			{Granell}}, \bibinfo {author} {\bibfnamefont {S.}~\bibnamefont {G\'omez}},\
		and\ \bibinfo {author} {\bibfnamefont {A.}~\bibnamefont {Arenas}},\
	}\bibfield  {title} {\bibinfo {title} {Dynamical interplay between awareness
			and epidemic spreading in multiplex networks},\ }\href
	{https://doi.org/10.1103/PhysRevLett.111.128701} {\bibfield  {journal}
		{\bibinfo  {journal} {Phys. Rev. Lett.}\ }\textbf {\bibinfo {volume} {111}},\
		\bibinfo {pages} {128701} (\bibinfo {year} {2013})}\BibitemShut {NoStop}%
	\bibitem [{\citenamefont {Wang}\ \emph {et~al.}(2016)\citenamefont {Wang},
		\citenamefont {Liu}, \citenamefont {Cai}, \citenamefont {Tang}, \citenamefont
		{Braunstein},\ and\ \citenamefont {Stanley}}]{Wang2016}%
	\BibitemOpen
	\bibfield  {author} {\bibinfo {author} {\bibfnamefont {W.}~\bibnamefont
			{Wang}}, \bibinfo {author} {\bibfnamefont {Q.-H.}\ \bibnamefont {Liu}},
		\bibinfo {author} {\bibfnamefont {S.-M.}\ \bibnamefont {Cai}}, \bibinfo
		{author} {\bibfnamefont {M.}~\bibnamefont {Tang}}, \bibinfo {author}
		{\bibfnamefont {L.~A.}\ \bibnamefont {Braunstein}},\ and\ \bibinfo {author}
		{\bibfnamefont {H.~E.}\ \bibnamefont {Stanley}},\ }\bibfield  {title}
	{\bibinfo {title} {Suppressing disease spreading by using information
			diffusion on multiplex networks},\ }\href {https://doi.org/10.1038/srep29259}
	{\bibfield  {journal} {\bibinfo  {journal} {Scientific Reports}\ }\textbf
		{\bibinfo {volume} {6}},\ \bibinfo {pages} {29259} (\bibinfo {year}
		{2016})}\BibitemShut {NoStop}%
	\bibitem [{\citenamefont {Bedson}\ \emph {et~al.}(2021)\citenamefont {Bedson},
		\citenamefont {Skrip}, \citenamefont {Pedi}, \citenamefont {Abramowitz},
		\citenamefont {Carter}, \citenamefont {Jalloh}, \citenamefont {Funk},
		\citenamefont {Gobat}, \citenamefont {Giles-Vernick}, \citenamefont
		{Chowell}, \citenamefont {de~Almeida}, \citenamefont {Elessawi},
		\citenamefont {Scarpino}, \citenamefont {Hammond}, \citenamefont {Briand},
		\citenamefont {Epstein}, \citenamefont {H\'ebert-Dufresne},\ and\
		\citenamefont {Althouse}}]{Bedson2021}%
	\BibitemOpen
	\bibfield  {author} {\bibinfo {author} {\bibfnamefont {J.}~\bibnamefont
			{Bedson}}, \bibinfo {author} {\bibfnamefont {L.~A.}\ \bibnamefont {Skrip}},
		\bibinfo {author} {\bibfnamefont {D.}~\bibnamefont {Pedi}}, \bibinfo {author}
		{\bibfnamefont {S.}~\bibnamefont {Abramowitz}}, \bibinfo {author}
		{\bibfnamefont {S.}~\bibnamefont {Carter}}, \bibinfo {author} {\bibfnamefont
			{M.~F.}\ \bibnamefont {Jalloh}}, \bibinfo {author} {\bibfnamefont
			{S.}~\bibnamefont {Funk}}, \bibinfo {author} {\bibfnamefont {N.}~\bibnamefont
			{Gobat}}, \bibinfo {author} {\bibfnamefont {T.}~\bibnamefont
			{Giles-Vernick}}, \bibinfo {author} {\bibfnamefont {G.}~\bibnamefont
			{Chowell}}, \bibinfo {author} {\bibfnamefont {J.~a.~R.}\ \bibnamefont
			{de~Almeida}}, \bibinfo {author} {\bibfnamefont {R.}~\bibnamefont
			{Elessawi}}, \bibinfo {author} {\bibfnamefont {S.~V.}\ \bibnamefont
			{Scarpino}}, \bibinfo {author} {\bibfnamefont {R.~A.}\ \bibnamefont
			{Hammond}}, \bibinfo {author} {\bibfnamefont {S.}~\bibnamefont {Briand}},
		\bibinfo {author} {\bibfnamefont {J.~M.}\ \bibnamefont {Epstein}}, \bibinfo
		{author} {\bibfnamefont {L.}~\bibnamefont {H\'ebert-Dufresne}},\ and\
		\bibinfo {author} {\bibfnamefont {B.~M.}\ \bibnamefont {Althouse}},\
	}\bibfield  {title} {\bibinfo {title} {A review and agenda for integrated
			disease models including social and behavioural factors},\ }\href
	{https://doi.org/10.1038/s41562-021-01136-2} {\bibfield  {journal} {\bibinfo
			{journal} {Nature Human Behaviour}\ }\textbf {\bibinfo {volume} {5}},\
		\bibinfo {pages} {834} (\bibinfo {year} {2021})}\BibitemShut {NoStop}%
	\bibitem [{\citenamefont {Chen}\ \emph {et~al.}(2013)\citenamefont {Chen},
		\citenamefont {Ghanbarnejad}, \citenamefont {Cai},\ and\ \citenamefont
		{Grassberger}}]{Chen2013}%
	\BibitemOpen
	\bibfield  {author} {\bibinfo {author} {\bibfnamefont {L.}~\bibnamefont
			{Chen}}, \bibinfo {author} {\bibfnamefont {F.}~\bibnamefont {Ghanbarnejad}},
		\bibinfo {author} {\bibfnamefont {W.}~\bibnamefont {Cai}},\ and\ \bibinfo
		{author} {\bibfnamefont {P.}~\bibnamefont {Grassberger}},\ }\bibfield
	{title} {\bibinfo {title} {{Outbreaks of coinfections: The critical role of
				cooperativity}},\ }\href
	{http://stacks.iop.org/0295-5075/104/i=5/a=50001?key=crossref.285529110af16862399b90e7a308d85d}
	{\bibfield  {journal} {\bibinfo  {journal} {EPL (Europhysics Lett.}\ }\textbf
		{\bibinfo {volume} {104}},\ \bibinfo {pages} {50001} (\bibinfo {year}
		{2013})}\BibitemShut {NoStop}%
	\bibitem [{\citenamefont {Cai}\ \emph {et~al.}(2015)\citenamefont {Cai},
		\citenamefont {Chen}, \citenamefont {Ghanbarnejad},\ and\ \citenamefont
		{Grassberger}}]{Cai2015}%
	\BibitemOpen
	\bibfield  {author} {\bibinfo {author} {\bibfnamefont {W.}~\bibnamefont
			{Cai}}, \bibinfo {author} {\bibfnamefont {L.}~\bibnamefont {Chen}}, \bibinfo
		{author} {\bibfnamefont {F.}~\bibnamefont {Ghanbarnejad}},\ and\ \bibinfo
		{author} {\bibfnamefont {P.}~\bibnamefont {Grassberger}},\ }\bibfield
	{title} {\bibinfo {title} {{Avalanche outbreaks emerging in cooperative
				contagions}},\ }\href {http://www.nature.com/articles/nphys3457} {\bibfield
		{journal} {\bibinfo  {journal} {Nat. Phys.}\ }\textbf {\bibinfo {volume}
			{11}},\ \bibinfo {pages} {936} (\bibinfo {year} {2015})}\BibitemShut
	{NoStop}%
	\bibitem [{\citenamefont {Grassberger}\ \emph {et~al.}(2016)\citenamefont
		{Grassberger}, \citenamefont {Chen}, \citenamefont {Ghanbarnejad},\ and\
		\citenamefont {Cai}}]{Grassberger2016}%
	\BibitemOpen
	\bibfield  {author} {\bibinfo {author} {\bibfnamefont {P.}~\bibnamefont
			{Grassberger}}, \bibinfo {author} {\bibfnamefont {L.}~\bibnamefont {Chen}},
		\bibinfo {author} {\bibfnamefont {F.}~\bibnamefont {Ghanbarnejad}},\ and\
		\bibinfo {author} {\bibfnamefont {W.}~\bibnamefont {Cai}},\ }\bibfield
	{title} {\bibinfo {title} {{Phase transitions in cooperative coinfections:
				Simulation results for networks and lattices}},\ }\href
	{https://link.aps.org/doi/10.1103/PhysRevE.93.042316} {\bibfield  {journal}
		{\bibinfo  {journal} {Phys. Rev. E}\ }\textbf {\bibinfo {volume} {93}},\
		\bibinfo {pages} {042316} (\bibinfo {year} {2016})}\BibitemShut {NoStop}%
	\bibitem [{\citenamefont {Janssen}\ and\ \citenamefont
		{Stenull}(2016)}]{Janssen2016}%
	\BibitemOpen
	\bibfield  {author} {\bibinfo {author} {\bibfnamefont {H.-K.}\ \bibnamefont
			{Janssen}}\ and\ \bibinfo {author} {\bibfnamefont {O.}~\bibnamefont
			{Stenull}},\ }\bibfield  {title} {\bibinfo {title} {{First-order phase
				transitions in outbreaks of co-infectious diseases and the extended general
				epidemic process}},\ }\href
	{http://stacks.iop.org/0295-5075/113/i=2/a=26005?key=crossref.9ce6e00f12af4f7b438b0897232357d9}
	{\bibfield  {journal} {\bibinfo  {journal} {EPL (Europhysics Lett.}\ }\textbf
		{\bibinfo {volume} {113}},\ \bibinfo {pages} {26005} (\bibinfo {year}
		{2016})}\BibitemShut {NoStop}%
	\bibitem [{\citenamefont {Cui}\ \emph {et~al.}(2017)\citenamefont {Cui},
		\citenamefont {Colaiori},\ and\ \citenamefont {Castellano}}]{Cui2017}%
	\BibitemOpen
	\bibfield  {author} {\bibinfo {author} {\bibfnamefont {P.-B.}\ \bibnamefont
			{Cui}}, \bibinfo {author} {\bibfnamefont {F.}~\bibnamefont {Colaiori}},\ and\
		\bibinfo {author} {\bibfnamefont {C.}~\bibnamefont {Castellano}},\ }\bibfield
	{title} {\bibinfo {title} {{Mutually cooperative epidemics on power-law
				networks}},\ }\href {http://link.aps.org/doi/10.1103/PhysRevE.96.022301}
	{\bibfield  {journal} {\bibinfo  {journal} {Phys. Rev. E}\ }\textbf {\bibinfo
			{volume} {96}},\ \bibinfo {pages} {022301} (\bibinfo {year}
		{2017})}\BibitemShut {NoStop}%
	\bibitem [{\citenamefont {Liu}\ \emph {et~al.}(2017)\citenamefont {Liu},
		\citenamefont {Wang}, \citenamefont {Tang}, \citenamefont {Zhou},\ and\
		\citenamefont {Lai}}]{Liu}%
	\BibitemOpen
	\bibfield  {author} {\bibinfo {author} {\bibfnamefont {Q.}~\bibnamefont
			{Liu}}, \bibinfo {author} {\bibfnamefont {W.}~\bibnamefont {Wang}}, \bibinfo
		{author} {\bibfnamefont {M.}~\bibnamefont {Tang}}, \bibinfo {author}
		{\bibfnamefont {T.}~\bibnamefont {Zhou}},\ and\ \bibinfo {author}
		{\bibfnamefont {Y.}~\bibnamefont {Lai}},\ }\bibfield  {title} {\bibinfo
		{title} {{Explosive spreading on complex networks: The role of synergy}},\
	}\href {http://link.aps.org/doi/10.1103/PhysRevE.95.042320} {\bibfield
		{journal} {\bibinfo  {journal} {Phys. Rev. E}\ }\textbf {\bibinfo {volume}
			{95}},\ \bibinfo {pages} {042320} (\bibinfo {year} {2017})}\BibitemShut
	{NoStop}%
	\bibitem [{\citenamefont {Baek}\ \emph {et~al.}(2019)\citenamefont {Baek},
		\citenamefont {Chung}, \citenamefont {Ha}, \citenamefont {Jeong},\ and\
		\citenamefont {Kim}}]{Baek2019}%
	\BibitemOpen
	\bibfield  {author} {\bibinfo {author} {\bibfnamefont {Y.}~\bibnamefont
			{Baek}}, \bibinfo {author} {\bibfnamefont {K.}~\bibnamefont {Chung}},
		\bibinfo {author} {\bibfnamefont {M.}~\bibnamefont {Ha}}, \bibinfo {author}
		{\bibfnamefont {H.}~\bibnamefont {Jeong}},\ and\ \bibinfo {author}
		{\bibfnamefont {D.}~\bibnamefont {Kim}},\ }\bibfield  {title} {\bibinfo
		{title} {{Role of hubs in the synergistic spread of behavior}},\ }\href
	{https://link.aps.org/doi/10.1103/PhysRevE.99.020301} {\bibfield  {journal}
		{\bibinfo  {journal} {Phys. Rev. E}\ }\textbf {\bibinfo {volume} {99}},\
		\bibinfo {pages} {020301} (\bibinfo {year} {2019})}\BibitemShut {NoStop}%
	\bibitem [{\citenamefont {de~Oliveira}\ \emph {et~al.}(2012)\citenamefont
		{de~Oliveira}, \citenamefont {Dos~Santos},\ and\ \citenamefont
		{Dickman}}]{deOliveira2012}%
	\BibitemOpen
	\bibfield  {author} {\bibinfo {author} {\bibfnamefont {M.~M.}\ \bibnamefont
			{de~Oliveira}}, \bibinfo {author} {\bibfnamefont {R.~V.}\ \bibnamefont
			{Dos~Santos}},\ and\ \bibinfo {author} {\bibfnamefont {R.}~\bibnamefont
			{Dickman}},\ }\bibfield  {title} {\bibinfo {title} {Symbiotic two-species
			contact process},\ }\href
	{https://link.aps.org/doi/10.1103/PhysRevE.86.011121} {\bibfield  {journal}
		{\bibinfo  {journal} {Phys. Rev. E}\ }\textbf {\bibinfo {volume} {86}},\
		\bibinfo {pages} {011121} (\bibinfo {year} {2012})}\BibitemShut {NoStop}%
	\bibitem [{\citenamefont {de~Oliveira}\ \emph {et~al.}(2019)\citenamefont
		{de~Oliveira}, \citenamefont {Alves},\ and\ \citenamefont
		{Ferreira}}]{deOliveira2019}%
	\BibitemOpen
	\bibfield  {author} {\bibinfo {author} {\bibfnamefont {M.~M.}\ \bibnamefont
			{de~Oliveira}}, \bibinfo {author} {\bibfnamefont {S.~G.}\ \bibnamefont
			{Alves}},\ and\ \bibinfo {author} {\bibfnamefont {S.~C.}\ \bibnamefont
			{Ferreira}},\ }\bibfield  {title} {\bibinfo {title} {Dynamical correlations
			and pairwise theory for the symbiotic contact process on networks},\ }\href
	{https://link.aps.org/doi/10.1103/PhysRevE.100.052302} {\bibfield  {journal}
		{\bibinfo  {journal} {Phys. Rev. E}\ }\textbf {\bibinfo {volume} {100}},\
		\bibinfo {pages} {052302} (\bibinfo {year} {2019})}\BibitemShut {NoStop}%
	\bibitem [{\citenamefont {Pastor-Satorras}\ \emph {et~al.}(2015)\citenamefont
		{Pastor-Satorras}, \citenamefont {Castellano}, \citenamefont {{Van
				Mieghem}},\ and\ \citenamefont {Vespignani}}]{Pastor-Satorras2015}%
	\BibitemOpen
	\bibfield  {author} {\bibinfo {author} {\bibfnamefont {R.}~\bibnamefont
			{Pastor-Satorras}}, \bibinfo {author} {\bibfnamefont {C.}~\bibnamefont
			{Castellano}}, \bibinfo {author} {\bibfnamefont {P.}~\bibnamefont {{Van
					Mieghem}}},\ and\ \bibinfo {author} {\bibfnamefont {A.}~\bibnamefont
			{Vespignani}},\ }\bibfield  {title} {\bibinfo {title} {{Epidemic processes in
				complex networks}},\ }\href
	{https://link.aps.org/doi/10.1103/RevModPhys.87.925} {\bibfield  {journal}
		{\bibinfo  {journal} {Rev. Mod. Phys.}\ }\textbf {\bibinfo {volume} {87}},\
		\bibinfo {pages} {925} (\bibinfo {year} {2015})}\BibitemShut {NoStop}%
	\bibitem [{\citenamefont {Castellano}\ \emph {et~al.}(2009)\citenamefont
		{Castellano}, \citenamefont {Fortunato},\ and\ \citenamefont
		{Loreto}}]{Castellano2009}%
	\BibitemOpen
	\bibfield  {author} {\bibinfo {author} {\bibfnamefont {C.}~\bibnamefont
			{Castellano}}, \bibinfo {author} {\bibfnamefont {S.}~\bibnamefont
			{Fortunato}},\ and\ \bibinfo {author} {\bibfnamefont {V.}~\bibnamefont
			{Loreto}},\ }\bibfield  {title} {\bibinfo {title} {{Statistical physics of
				social dynamics}},\ }\href
	{https://link.aps.org/doi/10.1103/RevModPhys.81.591} {\bibfield  {journal}
		{\bibinfo  {journal} {Rev. Mod. Phys.}\ }\textbf {\bibinfo {volume} {81}},\
		\bibinfo {pages} {591} (\bibinfo {year} {2009})}\BibitemShut {NoStop}%
	\bibitem [{\citenamefont {Marro}\ and\ \citenamefont
		{Dickman}(1999)}]{Marro2005}%
	\BibitemOpen
	\bibfield  {author} {\bibinfo {author} {\bibfnamefont {J.}~\bibnamefont
			{Marro}}\ and\ \bibinfo {author} {\bibfnamefont {R.}~\bibnamefont
			{Dickman}},\ }\href
	{https://www.cambridge.org/core/product/identifier/9780511524288/type/book}
	{\emph {\bibinfo {title} {{Nonequilibrium Phase Transitions in Lattice
					Models}}}}\ (\bibinfo  {publisher} {Cambridge University Press},\ \bibinfo
	{address} {Cambridge,UK},\ \bibinfo {year} {1999})\BibitemShut {NoStop}%
	\bibitem [{\citenamefont {Harris}(1974)}]{Harris1974}%
	\BibitemOpen
	\bibfield  {author} {\bibinfo {author} {\bibfnamefont {T.~E.}\ \bibnamefont
			{Harris}},\ }\bibfield  {title} {\bibinfo {title} {{Contact Interactions on a
				Lattice}},\ }\href {https://doi.org/10.1214/aop/1176996493} {\bibfield
		{journal} {\bibinfo  {journal} {The Annals of Probability}\ }\textbf
		{\bibinfo {volume} {2}},\ \bibinfo {pages} {969 } (\bibinfo {year}
		{1974})}\BibitemShut {NoStop}%
	\bibitem [{\citenamefont {Castellano}\ and\ \citenamefont
		{Pastor-Satorras}(2006)}]{Castellano2006}%
	\BibitemOpen
	\bibfield  {author} {\bibinfo {author} {\bibfnamefont {C.}~\bibnamefont
			{Castellano}}\ and\ \bibinfo {author} {\bibfnamefont {R.}~\bibnamefont
			{Pastor-Satorras}},\ }\bibfield  {title} {\bibinfo {title} {{Non-Mean-Field
				Behavior of the Contact Process on Scale-Free Networks}},\ }\href
	{http://link.aps.org/doi/10.1103/PhysRevLett.96.038701} {\bibfield  {journal}
		{\bibinfo  {journal} {Phys. Rev. Lett.}\ }\textbf {\bibinfo {volume} {96}},\
		\bibinfo {pages} {038701} (\bibinfo {year} {2006})}\BibitemShut {NoStop}%
	\bibitem [{\citenamefont {Albert}\ and\ \citenamefont
		{Barab{\'{a}}si}(2002)}]{Albert2002}%
	\BibitemOpen
	\bibfield  {author} {\bibinfo {author} {\bibfnamefont {R.}~\bibnamefont
			{Albert}}\ and\ \bibinfo {author} {\bibfnamefont {A.-L.}\ \bibnamefont
			{Barab{\'{a}}si}},\ }\bibfield  {title} {\bibinfo {title} {{Statistical
				mechanics of complex networks}},\ }\href
	{https://link.aps.org/doi/10.1103/RevModPhys.74.47} {\bibfield  {journal}
		{\bibinfo  {journal} {Rev. Mod. Phys.}\ }\textbf {\bibinfo {volume} {74}},\
		\bibinfo {pages} {47} (\bibinfo {year} {2002})}\BibitemShut {NoStop}%
	\bibitem [{\citenamefont {Ferreira}\ \emph
		{et~al.}(2011{\natexlab{a}})\citenamefont {Ferreira}, \citenamefont
		{Ferreira}, \citenamefont {Castellano},\ and\ \citenamefont
		{Pastor-Satorras}}]{Ferreira2011}%
	\BibitemOpen
	\bibfield  {author} {\bibinfo {author} {\bibfnamefont {S.~C.}\ \bibnamefont
			{Ferreira}}, \bibinfo {author} {\bibfnamefont {R.~S.}\ \bibnamefont
			{Ferreira}}, \bibinfo {author} {\bibfnamefont {C.}~\bibnamefont
			{Castellano}},\ and\ \bibinfo {author} {\bibfnamefont {R.}~\bibnamefont
			{Pastor-Satorras}},\ }\bibfield  {title} {\bibinfo {title} {Quasistationary
			simulations of the contact process on quenched networks},\ }\href
	{https://link.aps.org/doi/10.1103/PhysRevE.84.066102} {\bibfield  {journal}
		{\bibinfo  {journal} {Phys. Rev. E}\ }\textbf {\bibinfo {volume} {84}},\
		\bibinfo {pages} {066102} (\bibinfo {year} {2011}{\natexlab{a}})}\BibitemShut
	{NoStop}%
	\bibitem [{\citenamefont {Mata}\ \emph {et~al.}(2014)\citenamefont {Mata},
		\citenamefont {Ferreira},\ and\ \citenamefont {Ferreira}}]{Mata2014}%
	\BibitemOpen
	\bibfield  {author} {\bibinfo {author} {\bibfnamefont {A.~S.}\ \bibnamefont
			{Mata}}, \bibinfo {author} {\bibfnamefont {R.~S.}\ \bibnamefont {Ferreira}},\
		and\ \bibinfo {author} {\bibfnamefont {S.~C.}\ \bibnamefont {Ferreira}},\
	}\bibfield  {title} {\bibinfo {title} {{Heterogeneous pair-approximation for
				the contact process on complex networks}},\ }\href
	{http://stacks.iop.org/1367-2630/16/i=5/a=053006?key=crossref.b50b1b736851ea680f948bb1a6065508}
	{\bibfield  {journal} {\bibinfo  {journal} {New J. Phys.}\ }\textbf {\bibinfo
			{volume} {16}},\ \bibinfo {pages} {053006} (\bibinfo {year}
		{2014})}\BibitemShut {NoStop}%
	\bibitem [{\citenamefont {Ferreira}\ \emph {et~al.}(2016)\citenamefont
		{Ferreira}, \citenamefont {Sander},\ and\ \citenamefont
		{Pastor-Satorras}}]{Ferreira2016a}%
	\BibitemOpen
	\bibfield  {author} {\bibinfo {author} {\bibfnamefont {S.~C.}\ \bibnamefont
			{Ferreira}}, \bibinfo {author} {\bibfnamefont {R.~S.}\ \bibnamefont
			{Sander}},\ and\ \bibinfo {author} {\bibfnamefont {R.}~\bibnamefont
			{Pastor-Satorras}},\ }\bibfield  {title} {\bibinfo {title} {{Collective
				versus hub activation of epidemic phases on networks}},\ }\href
	{http://arxiv.org/abs/1512.00316
		https://link.aps.org/doi/10.1103/PhysRevE.93.032314} {\bibfield  {journal}
		{\bibinfo  {journal} {Phys. Rev. E}\ }\textbf {\bibinfo {volume} {93}},\
		\bibinfo {pages} {032314} (\bibinfo {year} {2016})}\BibitemShut {NoStop}%
	\bibitem [{\citenamefont {Chatterjee}\ and\ \citenamefont
		{Durrett}(2009)}]{Chatterjee2009}%
	\BibitemOpen
	\bibfield  {author} {\bibinfo {author} {\bibfnamefont {S.}~\bibnamefont
			{Chatterjee}}\ and\ \bibinfo {author} {\bibfnamefont {R.}~\bibnamefont
			{Durrett}},\ }\bibfield  {title} {\bibinfo {title} {{Contact processes on
				random graphs with power law degree distributions have critical value 0}},\
	}\href {http://dx.doi.org/10.1214/09-AOP471} {\bibfield  {journal} {\bibinfo
			{journal} {Ann. Probab.}\ }\textbf {\bibinfo {volume} {37}},\ \bibinfo
		{pages} {2332} (\bibinfo {year} {2009})}\BibitemShut {NoStop}%
	\bibitem [{\citenamefont {Bogu{\~{n}}{\'{a}}}\ \emph
		{et~al.}(2013)\citenamefont {Bogu{\~{n}}{\'{a}}}, \citenamefont
		{Castellano},\ and\ \citenamefont {Pastor-Satorras}}]{Boguna2013}%
	\BibitemOpen
	\bibfield  {author} {\bibinfo {author} {\bibfnamefont {M.}~\bibnamefont
			{Bogu{\~{n}}{\'{a}}}}, \bibinfo {author} {\bibfnamefont {C.}~\bibnamefont
			{Castellano}},\ and\ \bibinfo {author} {\bibfnamefont {R.}~\bibnamefont
			{Pastor-Satorras}},\ }\bibfield  {title} {\bibinfo {title} {{Nature of the
				Epidemic Threshold for the Susceptible-Infected-Susceptible Dynamics in
				Networks}},\ }\href {https://link.aps.org/doi/10.1103/PhysRevLett.111.068701}
	{\bibfield  {journal} {\bibinfo  {journal} {Phys. Rev. Lett.}\ }\textbf
		{\bibinfo {volume} {111}},\ \bibinfo {pages} {068701} (\bibinfo {year}
		{2013})}\BibitemShut {NoStop}%
	\bibitem [{\citenamefont {Sanz}\ \emph {et~al.}(2014)\citenamefont {Sanz},
		\citenamefont {Xia}, \citenamefont {Meloni},\ and\ \citenamefont
		{Moreno}}]{Sanz2014}%
	\BibitemOpen
	\bibfield  {author} {\bibinfo {author} {\bibfnamefont {J.}~\bibnamefont
			{Sanz}}, \bibinfo {author} {\bibfnamefont {C.-Y.}\ \bibnamefont {Xia}},
		\bibinfo {author} {\bibfnamefont {S.}~\bibnamefont {Meloni}},\ and\ \bibinfo
		{author} {\bibfnamefont {Y.}~\bibnamefont {Moreno}},\ }\bibfield  {title}
	{\bibinfo {title} {{Dynamics of Interacting Diseases}},\ }\href
	{https://link.aps.org/doi/10.1103/PhysRevX.4.041005} {\bibfield  {journal}
		{\bibinfo  {journal} {Phys. Rev. X}\ }\textbf {\bibinfo {volume} {4}},\
		\bibinfo {pages} {041005} (\bibinfo {year} {2014})}\BibitemShut {NoStop}%
	\bibitem [{\citenamefont {de~Oliveira}\ and\ \citenamefont
		{Dickman}(2014)}]{deOliveira2014}%
	\BibitemOpen
	\bibfield  {author} {\bibinfo {author} {\bibfnamefont {M.~M.}\ \bibnamefont
			{de~Oliveira}}\ and\ \bibinfo {author} {\bibfnamefont {R.}~\bibnamefont
			{Dickman}},\ }\bibfield  {title} {\bibinfo {title} {Phase diagram of the
			symbiotic two-species contact process},\ }\href
	{https://link.aps.org/doi/10.1103/PhysRevE.90.032120} {\bibfield  {journal}
		{\bibinfo  {journal} {Phys. Rev. E}\ }\textbf {\bibinfo {volume} {90}},\
		\bibinfo {pages} {032120} (\bibinfo {year} {2014})}\BibitemShut {NoStop}%
	\bibitem [{\citenamefont {Sampaio~Filho}\ \emph {et~al.}(2018)\citenamefont
		{Sampaio~Filho}, \citenamefont {dos Santos}, \citenamefont {Ara\'ujo},
		\citenamefont {Carmona}, \citenamefont {Moreira},\ and\ \citenamefont
		{Andrade}}]{Sampaio2018}%
	\BibitemOpen
	\bibfield  {author} {\bibinfo {author} {\bibfnamefont {C.~I.~N.}\
			\bibnamefont {Sampaio~Filho}}, \bibinfo {author} {\bibfnamefont {T.~B.}\
			\bibnamefont {dos Santos}}, \bibinfo {author} {\bibfnamefont {N.~A.~M.}\
			\bibnamefont {Ara\'ujo}}, \bibinfo {author} {\bibfnamefont {H.~A.}\
			\bibnamefont {Carmona}}, \bibinfo {author} {\bibfnamefont {A.~A.}\
			\bibnamefont {Moreira}},\ and\ \bibinfo {author} {\bibfnamefont {J.~S.}\
			\bibnamefont {Andrade}},\ }\bibfield  {title} {\bibinfo {title} {Symbiotic
			contact process: Phase transitions, hysteresis cycles, and bistability},\
	}\href {https://doi.org/10.1103/PhysRevE.98.062108} {\bibfield  {journal}
		{\bibinfo  {journal} {Phys. Rev. E}\ }\textbf {\bibinfo {volume} {98}},\
		\bibinfo {pages} {062108} (\bibinfo {year} {2018})}\BibitemShut {NoStop}%
	\bibitem [{\citenamefont {Juh{\'{a}}sz}\ \emph {et~al.}(2012)\citenamefont
		{Juh{\'{a}}sz}, \citenamefont {{\'{O}}dor}, \citenamefont {Castellano},\ and\
		\citenamefont {Mu{\~{n}}oz}}]{Juhasz2012}%
	\BibitemOpen
	\bibfield  {author} {\bibinfo {author} {\bibfnamefont {R.}~\bibnamefont
			{Juh{\'{a}}sz}}, \bibinfo {author} {\bibfnamefont {G.}~\bibnamefont
			{{\'{O}}dor}}, \bibinfo {author} {\bibfnamefont {C.}~\bibnamefont
			{Castellano}},\ and\ \bibinfo {author} {\bibfnamefont {M.~A.}\ \bibnamefont
			{Mu{\~{n}}oz}},\ }\bibfield  {title} {\bibinfo {title} {{Rare-region effects
				in the contact process on networks}},\ }\href
	{http://link.aps.org/doi/10.1103/PhysRevE.85.066125} {\bibfield  {journal}
		{\bibinfo  {journal} {Phys. Rev. E}\ }\textbf {\bibinfo {volume} {85}},\
		\bibinfo {pages} {066125} (\bibinfo {year} {2012})}\BibitemShut {NoStop}%
	\bibitem [{\citenamefont {Pastor-Satorras}\ and\ \citenamefont
		{Vespignani}(2001)}]{Pastor-Satorras2001}%
	\BibitemOpen
	\bibfield  {author} {\bibinfo {author} {\bibfnamefont {R.}~\bibnamefont
			{Pastor-Satorras}}\ and\ \bibinfo {author} {\bibfnamefont {A.}~\bibnamefont
			{Vespignani}},\ }\bibfield  {title} {\bibinfo {title} {{Epidemic Spreading in
				Scale-Free Networks}},\ }\href
	{https://link.aps.org/doi/10.1103/PhysRevLett.86.3200} {\bibfield  {journal}
		{\bibinfo  {journal} {Phys. Rev. Lett.}\ }\textbf {\bibinfo {volume} {86}},\
		\bibinfo {pages} {3200} (\bibinfo {year} {2001})}\BibitemShut {NoStop}%
	\bibitem [{\citenamefont {Barab{\'{a}}si}\ and\ \citenamefont
		{P{\'{o}}sfai}(2016)}]{barabasi2016network}%
	\BibitemOpen
	\bibfield  {author} {\bibinfo {author} {\bibfnamefont {L.}~\bibnamefont
			{Barab{\'{a}}si}}\ and\ \bibinfo {author} {\bibfnamefont {M.}~\bibnamefont
			{P{\'{o}}sfai}},\ }\href {http://barabasi.com/networksciencebook/} {\emph
		{\bibinfo {title} {{Network science}}}}\ (\bibinfo  {publisher} {Cambridge
		University Press},\ \bibinfo {address} {Cambridge},\ \bibinfo {year}
	{2016})\BibitemShut {NoStop}%
	\bibitem [{\citenamefont {Zwillinger}(2015)}]{zwillinger2007table}%
	\BibitemOpen
	\bibfield  {author} {\bibinfo {author} {\bibfnamefont {D.}~\bibnamefont
			{Zwillinger}},\ }\href {https://doi.org/10.1016/C2010-0-64839-5} {\emph
		{\bibinfo {title} {Table of Integrals, Series, and Products}}}\ (\bibinfo
	{publisher} {Elsevier},\ \bibinfo {year} {2015})\ pp.\ \bibinfo {pages}
	{1--1133}\BibitemShut {NoStop}%
	\bibitem [{\citenamefont {Bogu{\~{n}}{\'{a}}}\ \emph
		{et~al.}(2009)\citenamefont {Bogu{\~{n}}{\'{a}}}, \citenamefont
		{Castellano},\ and\ \citenamefont {Pastor-Satorras}}]{Boguna2009}%
	\BibitemOpen
	\bibfield  {author} {\bibinfo {author} {\bibfnamefont {M.}~\bibnamefont
			{Bogu{\~{n}}{\'{a}}}}, \bibinfo {author} {\bibfnamefont {C.}~\bibnamefont
			{Castellano}},\ and\ \bibinfo {author} {\bibfnamefont {R.}~\bibnamefont
			{Pastor-Satorras}},\ }\bibfield  {title} {\bibinfo {title} {{Langevin
				approach for the dynamics of the contact process on annealed scale-free
				networks}},\ }\href {http://link.aps.org/doi/10.1103/PhysRevE.79.036110}
	{\bibfield  {journal} {\bibinfo  {journal} {Phys. Rev. E}\ }\textbf {\bibinfo
			{volume} {79}},\ \bibinfo {pages} {036110} (\bibinfo {year}
		{2009})}\BibitemShut {NoStop}%
	\bibitem [{\citenamefont {Strogatz}(2018)}]{Strogatz2018}%
	\BibitemOpen
	\bibfield  {author} {\bibinfo {author} {\bibfnamefont {S.~H.}\ \bibnamefont
			{Strogatz}},\ }\href {https://doi.org/10.1201/9780429492563} {\emph {\bibinfo
			{title} {{Nonlinear Dynamics and Chaos}}}}\ (\bibinfo  {publisher} {CRC
		Press},\ \bibinfo {year} {2018})\BibitemShut {NoStop}%
	\bibitem [{\citenamefont {Ferreira}\ \emph
		{et~al.}(2011{\natexlab{b}})\citenamefont {Ferreira}, \citenamefont
		{Ferreira},\ and\ \citenamefont {Pastor-Satorras}}]{Ferreira2011a}%
	\BibitemOpen
	\bibfield  {author} {\bibinfo {author} {\bibfnamefont {S.~C.}\ \bibnamefont
			{Ferreira}}, \bibinfo {author} {\bibfnamefont {R.~S.}\ \bibnamefont
			{Ferreira}},\ and\ \bibinfo {author} {\bibfnamefont {R.}~\bibnamefont
			{Pastor-Satorras}},\ }\bibfield  {title} {\bibinfo {title} {{Quasistationary
				analysis of the contact process on annealed scale-free networks}},\ }\href
	{http://link.aps.org/doi/10.1103/PhysRevE.83.066113} {\bibfield  {journal}
		{\bibinfo  {journal} {Phys. Rev. E}\ }\textbf {\bibinfo {volume} {83}},\
		\bibinfo {pages} {066113} (\bibinfo {year} {2011}{\natexlab{b}})}\BibitemShut
	{NoStop}%
	\bibitem [{\citenamefont {Catanzaro}\ \emph {et~al.}(2005)\citenamefont
		{Catanzaro}, \citenamefont {Bogu{\~{n}}{\'{a}}},\ and\ \citenamefont
		{Pastor-Satorras}}]{Catanzaro2005}%
	\BibitemOpen
	\bibfield  {author} {\bibinfo {author} {\bibfnamefont {M.}~\bibnamefont
			{Catanzaro}}, \bibinfo {author} {\bibfnamefont {M.}~\bibnamefont
			{Bogu{\~{n}}{\'{a}}}},\ and\ \bibinfo {author} {\bibfnamefont
			{R.}~\bibnamefont {Pastor-Satorras}},\ }\bibfield  {title} {\bibinfo {title}
		{{Generation of uncorrelated random scale-free networks}},\ }\href
	{http://link.aps.org/doi/10.1103/PhysRevE.71.027103} {\bibfield  {journal}
		{\bibinfo  {journal} {Phys. Rev. E}\ }\textbf {\bibinfo {volume} {71}},\
		\bibinfo {pages} {027103} (\bibinfo {year} {2005})}\BibitemShut {NoStop}%
	\bibitem [{\citenamefont {Costa}\ and\ \citenamefont
		{Ferreira}(2021)}]{Costa2021}%
	\BibitemOpen
	\bibfield  {author} {\bibinfo {author} {\bibfnamefont {G.~S.}\ \bibnamefont
			{Costa}}\ and\ \bibinfo {author} {\bibfnamefont {S.~C.}\ \bibnamefont
			{Ferreira}},\ }\bibfield  {title} {\bibinfo {title} {{Simple quasistationary
				method for simulations of epidemic processes with localized states}},\ }\href
	{https://linkinghub.elsevier.com/retrieve/pii/S0010465521001582} {\bibfield
		{journal} {\bibinfo  {journal} {Comput. Phys. Commun.}\ }\textbf {\bibinfo
			{volume} {267}},\ \bibinfo {pages} {108046} (\bibinfo {year}
		{2021})}\BibitemShut {NoStop}%
	\bibitem [{\citenamefont {{De Oliveira}}\ and\ \citenamefont
		{Dickman}(2005)}]{DeOliveira2005}%
	\BibitemOpen
	\bibfield  {author} {\bibinfo {author} {\bibfnamefont {M.~M.}\ \bibnamefont
			{{De Oliveira}}}\ and\ \bibinfo {author} {\bibfnamefont {R.}~\bibnamefont
			{Dickman}},\ }\bibfield  {title} {\bibinfo {title} {{How to simulate the
				quasistationary state}},\ }\href
	{https://link.aps.org/doi/10.1103/PhysRevE.71.016129} {\bibfield  {journal}
		{\bibinfo  {journal} {Phys. Rev. E}\ }\textbf {\bibinfo {volume} {71}},\
		\bibinfo {pages} {016129} (\bibinfo {year} {2005})}\BibitemShut {NoStop}%
	\bibitem [{\citenamefont {de~Arruda}\ \emph {et~al.}(2017)\citenamefont
		{de~Arruda}, \citenamefont {Cozzo}, \citenamefont {Peixoto}, \citenamefont
		{Rodrigues},\ and\ \citenamefont {Moreno}}]{DeArruda2015a}%
	\BibitemOpen
	\bibfield  {author} {\bibinfo {author} {\bibfnamefont {G.~F.}\ \bibnamefont
			{de~Arruda}}, \bibinfo {author} {\bibfnamefont {E.}~\bibnamefont {Cozzo}},
		\bibinfo {author} {\bibfnamefont {T.~P.}\ \bibnamefont {Peixoto}}, \bibinfo
		{author} {\bibfnamefont {F.~A.}\ \bibnamefont {Rodrigues}},\ and\ \bibinfo
		{author} {\bibfnamefont {Y.}~\bibnamefont {Moreno}},\ }\bibfield  {title}
	{\bibinfo {title} {{Disease Localization in Multilayer Networks}},\ }\href
	{http://link.aps.org/doi/10.1103/PhysRevX.7.011014} {\bibfield  {journal}
		{\bibinfo  {journal} {Phys. Rev. X}\ }\textbf {\bibinfo {volume} {7}},\
		\bibinfo {pages} {011014} (\bibinfo {year} {2017})}\BibitemShut {NoStop}%
	\bibitem [{\citenamefont {Sander}\ \emph {et~al.}(2016)\citenamefont {Sander},
		\citenamefont {Costa},\ and\ \citenamefont {Ferreira}}]{Sander2016}%
	\BibitemOpen
	\bibfield  {author} {\bibinfo {author} {\bibfnamefont {R.~S.}\ \bibnamefont
			{Sander}}, \bibinfo {author} {\bibfnamefont {G.~S.}\ \bibnamefont {Costa}},\
		and\ \bibinfo {author} {\bibfnamefont {S.~C.}\ \bibnamefont {Ferreira}},\
	}\bibfield  {title} {\bibinfo {title} {{Sampling methods for the
				quasistationary regime of epidemic processes on regular and complex
				networks}},\ }\href {http://link.aps.org/doi/10.1103/PhysRevE.94.042308}
	{\bibfield  {journal} {\bibinfo  {journal} {Phys. Rev. E}\ }\textbf {\bibinfo
			{volume} {94}},\ \bibinfo {pages} {042308} (\bibinfo {year}
		{2016})}\BibitemShut {NoStop}%
\end{thebibliography}

%apsrev4-2.bst 2019-01-14 (MD) hand-edited version of apsrev4-1.bst
%Control: key (0)
%Control: author (8) initials jnrlst
%Control: editor formatted (1) identically to author
%Control: production of article title (0) allowed
%Control: page (0) single
%Control: year (1) truncated
%Control: production of eprint (0) enabled
%

\end{document}